\documentclass[twocolumn]{openjournal}

\DeclareUnicodeCharacter{2212}{-}

\usepackage{amsmath}
\usepackage{booktabs}
\usepackage{lipsum}
\usepackage[T1]{fontenc}
\usepackage[breaklinks,colorlinks,citecolor=blue,urlcolor=blue,linkcolor=blue]{hyperref}
\usepackage{cleveref}
\usepackage{orcidlink}

\usepackage{savesym}
\savesymbol{tablenum}
\usepackage{siunitx}
\restoresymbol{SIX}{tablenum}

\newcommand{\Msun}{M_{\odot}}

\newcommand{\kms}{\mbox{${\rm km~s^{-1}}$}}

\usepackage{graphicx} % Required for inserting images

\begin{document}

\title{Cosmography of the Sloan Basin of Attraction and Neighborhood}

\author{\vspace{-1.3cm}
Daniel Pomar\`ede\,\orcidlink{0000-0003-2038-0488}$^{1,\star}$,
R. Brent Tully\,\orcidlink{0000-0002-9291-1981}$^{2}$,
Aur\'elien Valade\,\orcidlink{0009-0002-5203-5128}$^{3}$,
Noam Libeskind\,\orcidlink{0000-0002-6406-0016}$^{4}$,
and Yehuda Hoffman\,\orcidlink{0000-0002-8158-0566}$^{5}$\\
}

\affiliation{
$^{1}$Institut de Recherche sur les Lois Fondamentales de l’Univers, CEA, Universit\'e Paris-Saclay, 91191 Gif-sur-Yvette, France\\
$^{2}$Institute for Astronomy, University of Hawaii, 2680 Woodlawn Drive, Honolulu, HI 96822, USA\\
$^{3}$Aix Marseille Universit\'e, CNRS/IN2P3, CPPM, Marseille, France\\
$^{4}$Leibniz Institut f\"ur Astrophysik Potsdam (AIP), An der Sternwarte 16, D-144 Potsdam, Germany\\
$^{5}$Racah Institute of Physics, Hebrew University, Jerusalem 91904, Israel
}
\thanks{$^{\star}$E-mail:pomarede@cea.fr}

%\author[0000-0003-2038-0488]{Daniel Pomar\`ede}
%\affiliation{Institut de Recherche sur les Lois Fondamentales de l’Univers, CEA, Universit\'e Paris-Saclay, 91191 Gif-sur-Yvette, France}

%\author[0000-0002-9291-1981]{R. Brent Tully}
%\affiliation{Institute for Astronomy, University of Hawaii, 2680 Woodlawn Drive, Honolulu, HI 96822, USA}

%\author[0009-0002-5203-5128]{Aur\'elien Valade}
%\affiliation{Aix Marseille Université, CNRS/IN2P3, CPPM, Marseille, France}

%\author[0000-0002-6406-0016]{Noam Libeskind}
%\affiliation{Leibniz Institut f\"ur Astrophysik Potsdam (AIP), An der Sternwarte 16, D-144 Potsdam, Germany}

%\author[0000-0002-8158-0566]{Yehuda Hoffman}
%\affiliation{Racah Institute of Physics, Hebrew University, Jerusalem 91904, Israel}

%%%%%%%%%%%%%%%%%%%%%%%%%%%%%%%%%%%%%%%%%%%%%%%%%%%%%%%%%%%%%%%%%%%%%%%%%%%%%%%%%%%%%%%%%%%%%%%%%%%%%%%%%%%%%%

\begin{abstract}

The Sloan Great Wall is a dominant structure that is relatively nearby.
As well as evident in redshift survey maps, its presence is manifested in distortions to cosmic expansion.
Here, Hamiltonian Monte Carlo forward reconstruction in a $\Lambda$CDM framework gives probabilistic density and velocity fields constrained by the Cosmicflows-4 compendium of galaxy distances and radial velocities.
Streamlines of the reconstructed velocity field  started from arbitrary points in space can be followed to sinks, i.e. the minima of the gravitational potential, due to the distribution of mass.
A basin of attraction encompasses the volume of all streamlines ending at the same sink.
The solution can be assigned probabilities, with uncertainties associated with the imperfect data and the random nature of the $\Lambda$CDM model.

The Sloan basin of attraction is by far the largest basin in the study region, extending across a diameter of $\sim0.13c$.
It can be described by velocity streamlines that converge on the Sloan Great Wall, by the reconstructed density field, and by the network of filaments of the V-web, formulated by shear in the velocity field.
The discussion of these elements is augmented by a video and interactive models.
It is of interest to see the relationship of the Ho`oleilana baryon acoustic oscillation feature with the Sloan basin of attraction.

\end{abstract}

\section{Introduction}

The intent of this article is to explore the large scale structure of the Universe by means of cosmography. This mapping of the cosmic structure is carried out within 
the standard cosmological framework of the  $\Lambda$CDM model, the standard model of cosmology invoking dark matter and dark energy within a cosmos that is homogeneous and isotropic on large scales.
The evident local departures from homogeneity and isotropy are treated as perturbations imprinted by quantum fluctuations at the epoch of the cosmic inflation.
However there are known ``tensions" with the standard model, among them the uncertain values of the Hubble constant, the structure growth parameter $f\sigma_8$ and, of particular interest with this study, the bulk flow.
%\bf{(YH: There is no tension with respect to the bulk flow. The CF4 bulk flow is consistent with LCDM to with (1-1.5)$\sigma$ on all scales - both by the BGC and the HMC reconstructions.)}}
The historical background is covered in the review edited by \citet{2024divalentino} and two personal recollections by \citet{2020coce.book.....P} and \citet{2024einasto}.

The standard methods of cosmology research involve the analysis of the large scale structure of the Universe by statistical tools aimed at the extraction of a few cosmological parameters (e.g. Hubble's constant, matter density, etc.) and/or simple functions (such as correlation functions and the  power spectrum) from the wealth of cosmological observational data.
Elaborate and sophisticated tools have been designed and  applied to that data, leading  eventually to the formulation of  the $\Lambda$CDM  model. Yet, the road leading to that analysis started with the study of the cosmography of the observed universe, the mapping of structure without theoretical preconceptions.
Given the wealth of new data that are now available on the large scale distribution of galaxies and their velocities we wish here to resurrect the tool of cosmography and to try and shed new light on the issue of how light and mass are distributed in our local Universe, hoping that such a study will enable a fresh new theoretical insight into cosmology.

There was a seminal period circa 1980 when evidence was mounting for the existence of some form of dark matter and theorists debated top-down and bottom-up structure formation possibilities.  Statistical tools such as the two-point correlation function only partially described clustering.
The revelation of the Universe of filaments and voids came from the 3D mapping of galaxy positions from redshift surveys over contiguous volumes \citep{1978MNRAS.185..357J, 1978ApJ...222..784G, 1982ApJ...257..389T, 1985AJ.....90.2445G, 1986ApJ...302L...1D}.
The evidence from simple maps was unambiguous.
It soon followed that similar patterns were found in n-body simulations \citep{1985ApJ...292..371D}.
The way was opened for a fruitful interplay between observations and theory.

This story needs to step back to recognize the pioneers.
\citet{1937HarCi.423....1S} came to appreciate that there were regions of the sky and, from apparent magnitudes, in distance where there were abundances of galaxies far above the average.
A particularly prominent one is now known as the Shapley concentration \citep{1989Natur.342..251R, 1989Natur.338..562S}.
As photographic coverage of the north celestial sky became available with the Palomar Sky Survey, \citet{1958ApJS....3..211A} and \citet{1961cgcg.book.....Z, 1968cgcg.bookR....Z} cataloged the occurrences of rich clusters of galaxies and regions of enhanced galaxy densities.
The Abell inventory has subsequently been extended all-sky \citep{1989ApJS...70....1A}.

On a more local scale, the authority was de Vaucouleurs.
He provided a rudimentary description of what he called the Local (Virgo) Supercluster \citep{1958Natur.182.1478D}.
Nearby galaxies are strongly concentrated toward a plane that de Vaucouleurs used to define the equator of the supergalactic coordinate system.
By happenstance, the equatorial plane passes close (within $6.32^{\circ}$) to the north and south poles in Galactic coordinates.
As a consequence, whereas in Galactic coordinates the equatorial band is obscured and regions of extragalactic interest are at the poles, in supergalactic coordinates the minimally obscured regions are in two lobes on the equator with the Milky Way plane at the poles.
In cartesian coordinates with SGX=0 at one of the two intersections of the Galactic and supergalactic planes and SGZ normal to the supergalactic plane, by construction SGY$>0$ is north of the Milky Way (except for an obscured wedge from the $6.32^{\circ}$ tilt).

This attention to the supergalactic coordinate system is motivated by its usefulness locally \citep[e.g.][]{2015MNRAS.452.1052L} but also far beyond the volume studied by de Vaucouleurs.
Especially to be noted is the alignment of large scale flows with the supergalactic equator, documented from early velocity measurements toward the ``great attractor" \citep{1987ApJ...313L..37D} to most recent reconstructions from the CF4 data  \citep{2024MNRAS.527.3788H,2024NatAs.tmp..234V}.
De Vaucouleurs did not have this information.

Redshift surveys provide the most abundant information on large scale structure.
The cosmic web \citep{1996Natur.380..603B} network of clusters, filaments, sheets, and voids can be discerned assuming that the luminosity associated with galaxies is correlated with mass.
An industry in astronomy involves statistical comparisons between the redshift structure mapped in galaxies with simulations, whether they only incorporate collisionless dark matter \citep{1985ApJ...292..371D} or involve increasingly realistic hydrodynamics \citep{2018MNRAS.473.1195L, 2018MNRAS.473.4077P, 2023MNRAS.526.4978S}.
A particular topic of interest relates to the sizes of the largest structures, both at high densities (clusters, filaments) or low densities (voids).  From redshift surveys alone, one can infer the locations and morphologies of prominent structures.
Following de Vaucouleurs, contiguous high density regions have been called superclusters, although the threshold characteristics of the term supercluster are ambiguous.
Various statistical measures, Minkowski functionals \citep{1994A&A...288..697M}, shapefinders and derivative genus topology, planarity, and filamentarity specifications \citep{1998ApJ...495L...5S},  have been developed that provide comparisons between observations and simulations \citep{2023MNRAS.521.4712B}.
Overall there is acceptable agreement between observations and expectations from simulations assuming the standard $\Lambda$CDM cosmology, although the largest observed structures are surprisingly large \citep{2023MNRAS.519.3227S} and there are hints of issues with the $f\sigma_8$ parameter monitoring the growth of structure with time \citep{2024MNRAS.531..788H}.

Measurement of galaxy distances independent of redshifts allow observed velocities to be separated into components of cosmic expansion and deviations called peculiar velocities.
These departures are attributed to the distribution of mass.  With redshift information alone there is ambiguity between light and mass, including the possibility of missing components, and issues of redshift-space distortions.

The down-side in a study involving peculiar velocities comes from the large uncertainties in individual distance measurements.
A 20\% error at $z=0.05$ translates to a 3,000~\kms\ uncertainty!  Evidently, a very large number of distance measurements are required to tease out meaningful results.

The Cosmicflows-4 (CF4) collection of 56,000 distances \citep{2023ApJ...944...94T} is used in this study; see also \citet{2023A&A...678A.176D}.  
Here, the analysis that leads to 3D maps of the velocity and density fields is carried out with the Hamiltonian forward modeling algorithm HAMLET \citep{2022MNRAS.513.5148V, 2023MNRAS.519.2981V} as implemented with CF4 data in the mapping of structure out to $z=0.1$ by \citet{2024NatAs.tmp..234V}.
In that study, the focus of attention is on the volume at $z<0.05$ and, in particular, on the probability of the association of the Milky Way with a basin of gravitational attraction either centered in Ophiuchus at $z\sim0.03$ or in the Shapley concentration at $z\sim0.05$. 
It was noted without much detail that the largest basin of attraction in the CF4 survey volume is associated with the Sloan Great Wall.
It is that structure which is given attention here.

\smallskip
\section{A natural division within the CF4 volume at 14,000~\kms.}

Within $z\sim0.05$, CF4 provides reasonable full sky coverage (aside from the zone of obscuration) from large Fundamental Plane (FP) and Tully-Fisher (TF) samples but at $0.05<z<0.1$ the only extensive coverage is provided by the Sloan Digital Sky Survey Peculiar Velocity catalog \citep{2022MNRAS.515..953H} with a distribution in space and velocity illustrated in Figs. 8 and 9 in \citet{2023ApJ...944...94T}.
This extension is in the north Galactic polar cap in the north celestial hemisphere.
It is of great convenience for the present discussion that voids create a natural near-far separation between major high density structures in the Sloan survey region: the Center for Astrophysics (CfA) Great Wall \citep{1986ApJ...302L...1D} and Hercules complex \citep{1979ApJ...234..793T} to the foreground at 6,000$-$11,000~\kms\ and the Sloan Great Wall \citep{2004ogci.conf....5V, 2005ApJ...624..463G} and related structures to the background at $\sim25,000$~\kms.

Much of the cosmography of the Sloan Great Wall region has been carried out by the group at Tartu Observatory making use of incremental releases of the Sloan Digital Sky Survey.
Their publications provide a framework for the current discussion through their identification of the most important cluster and supercluster structures \citep{2003A&A...405..425E, 2010A&A...522A..92E, 2012A&A...539A..80L, 2014MNRAS.438.3465T}.
They have given particular attention to the two highest luminosity peaks: the supercluster associated with Abell\,2142 \citep{2015A&A...580A..69E, 2018A&A...620A.149E, 2020A&A...641A.172E} and the Corona Borealis supercluster \citep{2021A&A...649A..51E}.
They describe the Sloan Great Wall as a complex of superclusters with collapsing cores \citep{2016A&A...595A..70E}.
These structures will be given attention within the context of the current analysis.

%%%%%%%%%%%%%%%%%%%
\section{Overview of the Hamiltonian Monte Carlo Analysis of Cosmicflows-4 Distances.}

\begin{figure}
    \centering
    \includegraphics[width=1.\linewidth]{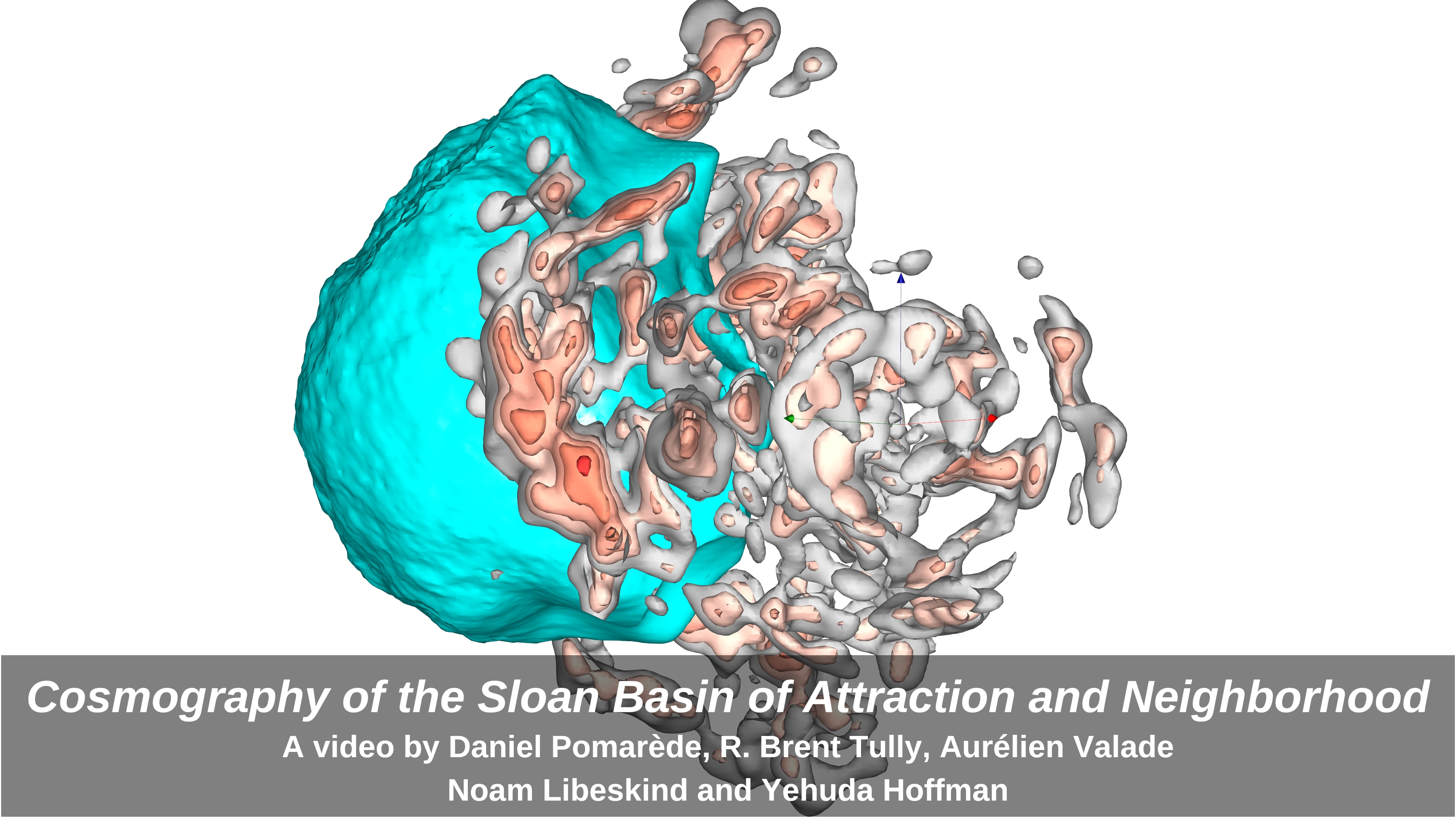}
    \caption{\href{https://vimeo.com/1073566900/851d5220d2}{Video visualization of the cosmography of the Sloan Basin of Attraction and neighborhood} $\leftarrow$ click on this link.
    }
    \label{fig:video}
\end{figure}

The following discussion will be coordinated with a video, accessed through Figure~\ref{fig:video}, and interludes when the reader will be invited to open interactive models.
To begin, there is a review of material presented by \citet{2024NatAs.tmp..234V}.
In brief, the 56,000 distance measurements in CF4, condensed into 38,000 groups \citep{2023ApJ...944...94T}, are translated into line-of-sight peculiar velocities, $V_{pec} \sim V_{obs} - H_0 d$, and assuming linear theory, infer 3D velocity and density fields from the Bayesian forward modeling Hamiltonian Monte Carlo (HMC) analysis.
Technical details on the modeling are discussed by \citet{2022MNRAS.513.5148V, 2023MNRAS.519.2981V}.
The method efficiently constructs a large number of solutions (here 1000) for the velocity and density fields consistent with $\Lambda$CDM cosmology and the CF4 data.
Averages over the solutions give mean density and velocity fields and variations in the velocity flow patterns provide statistical inferences on the reality of basins of attraction (BoA), the volumes associated with sinks of streamlines.
Probable basins of attraction, $p$-BoA, define volumes associated with a given sink region with probability fraction $p$.

%%%%%%%%%%%%%%%%%%%%%%%%%%%%%%%%%%%%%%%
%%%%%%%%%%%%%%%%%%%%%%%%%%%%%%%%%%%%%%%

%%%%%%%%%%%%%%%%%%%%%%%%%%%%%%%%%%%%%%%
\begin{figure}
    \centering
    \includegraphics[width=1.\linewidth]{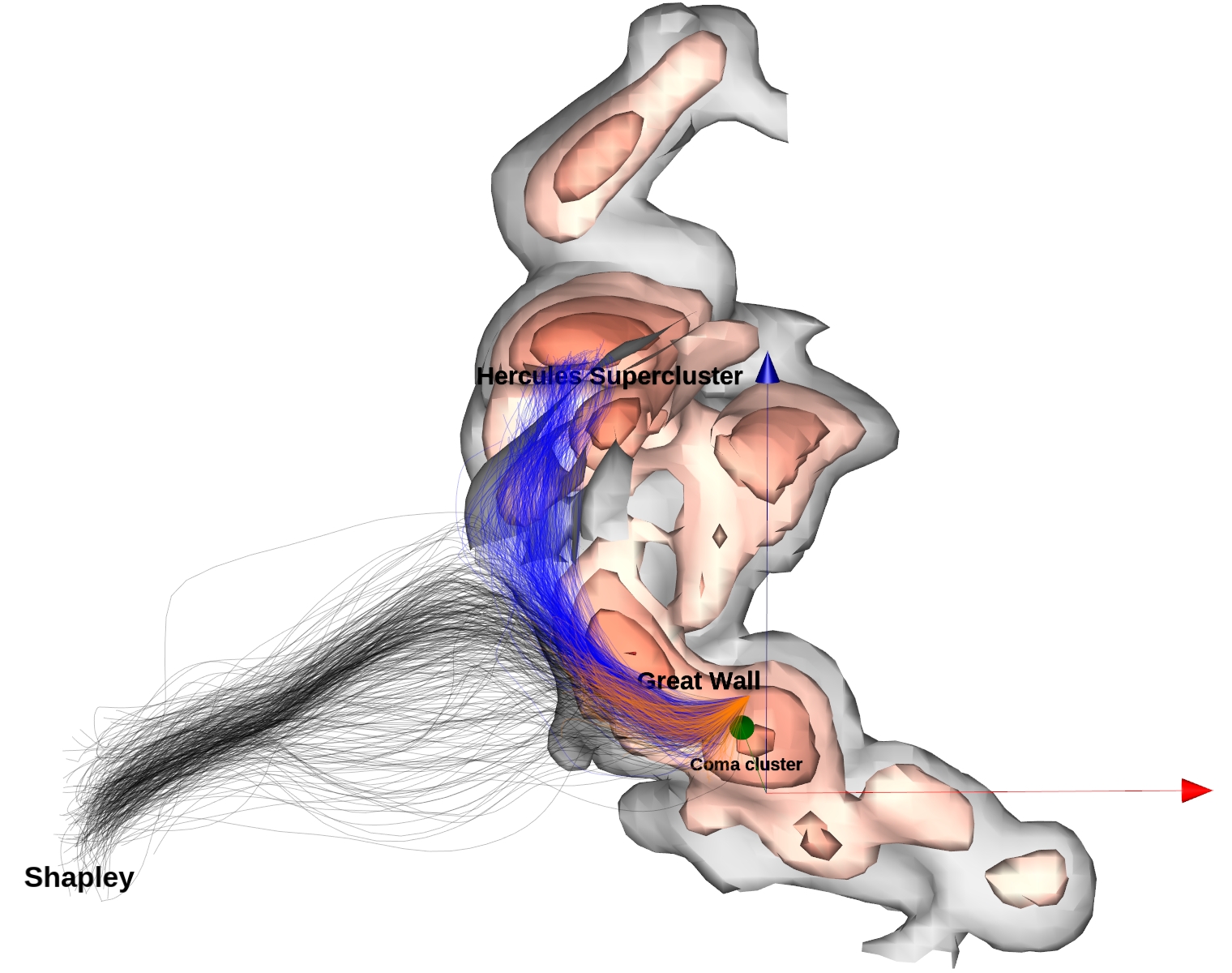}
    \caption{Contours of over density involving the CfA Great Wall and Hercules complex. Velocity streamlines from 1000 trials emanate from the Coma Cluster as a source. Orange streamlines illustrate the 25\% that terminate within the CfA Great Wall, the blue streamlines illustrate the 1/3 that terminate in the Hercules region, and the black lines illustrate the 40\% that terminate in the Shapley concentration. Scene at 01:45 in video and in this \href{https://sketchfab.com/3d-models/individual-realizations-seeded-near-coma-cluster-67b255a3fa14427ab1176ebec33f0070}{interactive visualization}.
    }
    \label{fig:coma}
\end{figure}
%%%%%%%%%%%%%%%%%%%%%%%%%%%%%%%%%%%%%%%

The video begins with a visualization of the CF4 data in supergalactic coordinates.  The asymmetry at SGY$\ge$14,000~\kms\ will become the focus of discussion in later sections.
Then briefly there are examples of individual HMC velocity-density solutions, followed by a tour of the mean density field.
The CfA Great Wall and Hercules complex are nearer than the volume of our primary interest but at the boundary so the video zooms to isolate these over dense features.

In \citet{2024NatAs.tmp..234V} there was the demonstration of the $p$-BoA analysis sourced at the Milky Way.
With $\sim60\%$ probability, the sinks of streamlines from our position end in the Shapley BoA.
With $\sim40\%$ probability they end in what was called the Ophiuchus BoA, an enlarged Laniakea Supercluster \citep{2014Natur.513...71T}.
In the video here, we provide an equivalent demonstration sourced at the Coma Cluster.
It is found that streamlines find sinks associated with the Shapley BoA 40\% of the time, with the Hercules BoA 1/3 of the time, and stand alone as a CfA Great Wall BoA 1/4 of the time.
Contoured high density surfaces of the CfA Great Wall$-$Hercules$-$Shapley features are shown in the first interactive model in Figure~\ref{fig:coma}.

%%%%%%%%%%%%%%%%%%%
\section{Overview of the Sloan Great Wall Neighborhood.}

Immediately beyond the CfA Great Wall$-$Hercules structures lie an interconnected complex of voids that create a natural separation from the over densities of the Sloan Great Wall region.  The video displays a schematic of the most pronounced void, seen at the mean HMC density level $\delta=-0.7$ to include as a part of it the early discovered Bo\"otes Void \citep{1987ApJ...314..493K}.
This encompassing void remarkably mirrors the adjacent Sloan Great Wall (SGW) and for that reason we name it the Sloan Great Void (SGV), see video at 02:45.

The ensuing discussion is restricted to the north Galactic pole cap at SGY$>$14,000~\kms, the domain of major overdensities behind the complex of voids.
Overwhelmingly, the data in CF4 that inform the HMC modeling of the velocity and density fields are from the Sloan Digital Sky Survey (SDSS) catalog of peculiar velocities (PV) from Fundamental Plane (FP) measurements \citep{2022MNRAS.515..953H}.
There is an inconsequential supplement of type\,Ia supernova distances. 

%%%%%%%%%%%%%%%%%%%%%%%%%%%%%%%%%%%%%%%
\begin{figure}
    \centering
    \includegraphics[width=1.\linewidth]{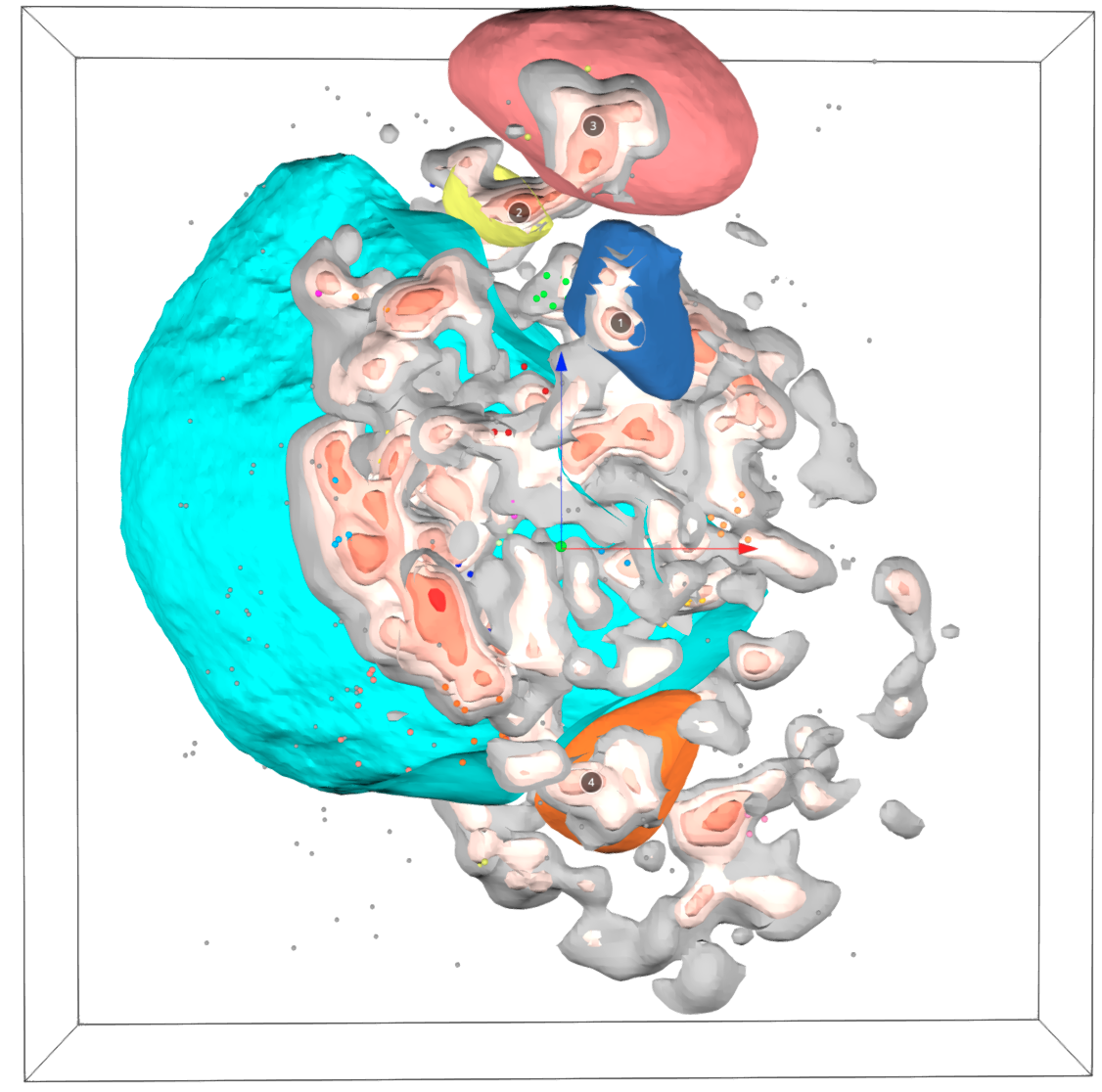}
    \caption{Shells of basins of attraction at the probability level p=0.5 for five major BoA at SGY$>$14,000~\kms\ at north Galactic latitudes superimposed on isocontours of high density from the mean HMC model.  The shell of the Sloan BoA is colored cyan, Corona~Borealis+19.6 is blue, Serpans~Caput+26.2 is yellow, Hercules+28.6 is red, and Leo+23.2 is tan. Scene at 03:17 in video.
    View it in 3D in this \href{https://sketchfab.com/3d-models/sloan-great-wall-208c8c497c12466685fd2848492e2a50}{interactive visualization}.}
    \label{fig:p0.5}
\end{figure}
%%%%%%%%%%%%%%%%%%%%%%%%%%%%%%%%%%%%%%%

The major basins of attraction in this region are identified  at probability $p=0.5$ in Figure~\ref{fig:p0.5} (structures near the observational boundaries will not be discussed).
The Sloan BoA is by far the largest.  
Four BoA besides Sloan are identified in the figure.
Following our convention, BoA are named after the constellation they lie in plus their systemic velocity in units of 1000~\kms\ (the convention is the same for voids except the sign preceding the velocity is negative).

At 03:50 in the video separate isocontours containing the Sloan BoA are shown at probability levels 0.25, 0.5, and 0.75.  The acquisition of regions of high density at probabilities below $p=0.5$ is modest.  Likewise, it will be shown later that the boundaries of the smaller BoA in the Sloan region are relatively stable with respect to variations in probability $p$.
Accordingly, the discussion in the next section of the properties of the Sloan BoA will be of the entity within $p=0.5$.

%%%%%%%%%%%%%%%%%%%%%%%%%%%%%%%%%%%%%%
\section{The Sloan Basin of Attraction}

A most evident fact is that the Sloan BoA fills a large part of the CF4 volume in the north Galactic - north celestial sector between $14,000-30,000$~\kms.
The data constraints are reasonable around most of the $p=0.5$ isosurface but {\it not} at the boundary set by the SDSS southern limit of $\delta=-3.7^{\circ}$. 
The Two Degree Field Galactic Redshift Survey was used to push south only to $\delta=-5.5^{\circ}$ \citep{2008ApJ...685...83E}.
The Las Campanus Redshift Survey provides a fragmented view with slices at $\delta=-6^{\circ}$ and $-12^{\circ}$ \citep{2003A&A...405..821E}.
The Six Degree Field Galaxy Survey \citep{2009MNRAS.399..683J} looses coverage beyond $z\sim 0.06$.
The current information available is too limited to bound the Sloan BoA to the celestial south.

A useful place to start a discussion is with the "superclusters" defined from friends-of-friends groupings of Abell \citep{1958ApJS....3..211A, 1989ApJS...70....1A} ACO and x-ray clusters by \citet{1994MNRAS.269..301E, 2001AJ....122.2222E}.  There is also the supercluster catalog by \citet{2012A&A...539A..80L} with an alternative naming convention that will not be used.
\citet{2024A&A...681A..91E} discuss the overall properties of groups and clusters in the SDSS main spectroscopic survey region.

%%%%%%%%%%%%%%%%%%%%%%%%%%%%%%%%%%%%%%%
\begin{figure}
    \centering
    \includegraphics[width=1.\linewidth]{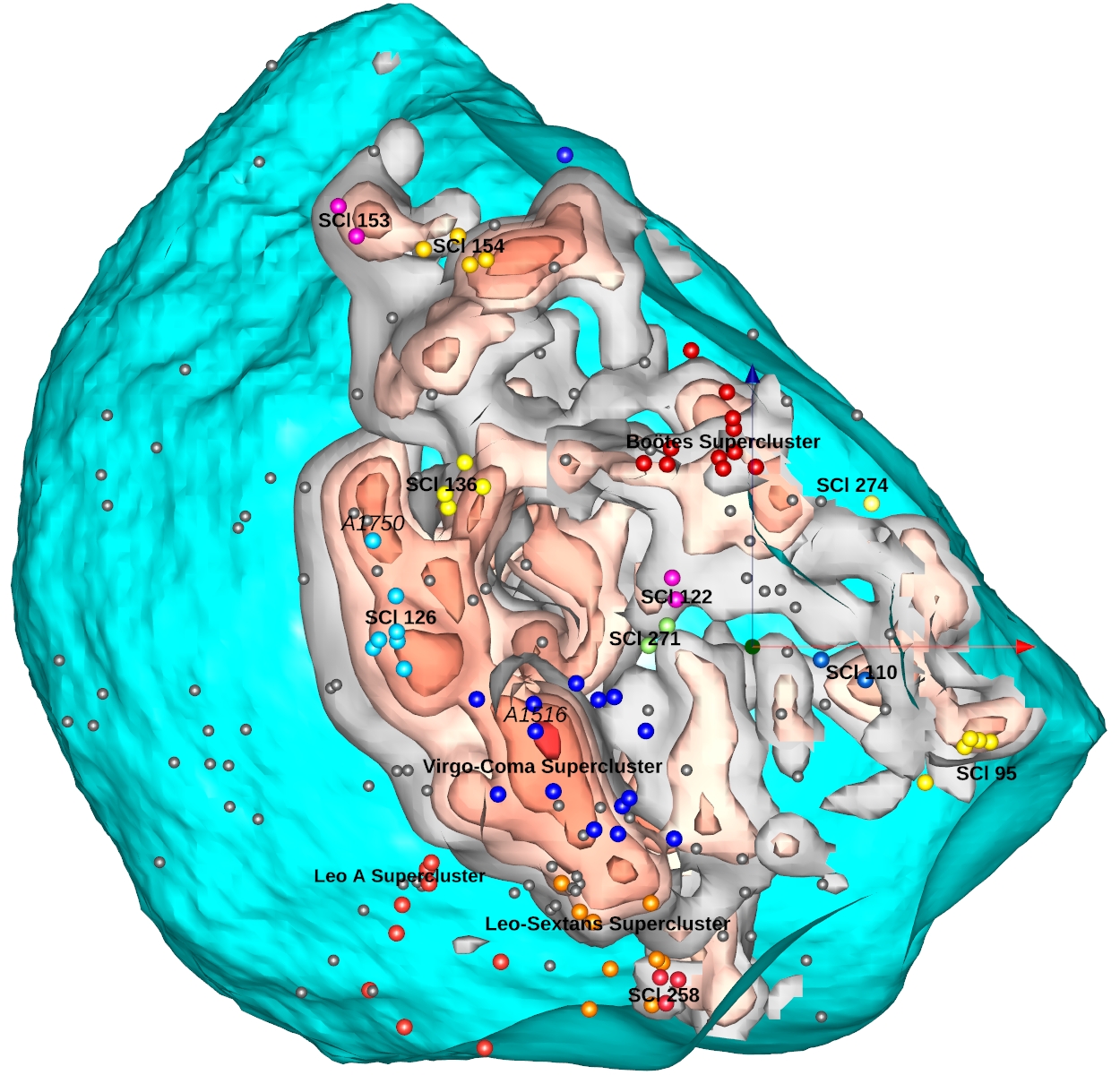}
    \caption{The $p=0.5$ shell of the Sloan BoA, the average of 1000 HMC realizations, with overdensity contours from the mean HMC density field at $\delta$ 1.2,1.7,2.2,2.8 (levels used in subsequent figures).  ACO clusters within the Sloan BoA are located by open circles. Distinct colors are given to those in 14 superclusters given names in the figure \citep{1994MNRAS.269..301E}.
    Scene at 04:25 in video.
    View it in 3D in this \href{https://sketchfab.com/3d-models/sloan-basin-of-attraction-6ce57733c89c4ffca92e40fe7f48c870}{interactive visualization}.}
    \label{fig:sloanscl}
\end{figure}
%%%%%%%%%%%%%%%%%%%%%%%%%%%%%%%%%%%%%%%

At 04:25 in the video the locations of ACO clusters are added to the scene, with distinctive colors associated with named superclusters.  The five dominant structures (with numbers of ACO clusters/x-ray clusters in brackets) are Virgo-Coma = Scl\,111 (15/5), Bo\"otes = Scl\,138 (12/6), Leo-Sextans = Scl\,91 (9/1), Leo\,A = Scl\,100 (9/0), and Scl\,126 (7/0).  
The core of what has been called the Sloan Great Wall \citep{2005ApJ...624..463G} arises from the union of Scl\,111 (Virgo-Coma) and Scl\,126 \citep{2010A&A...522A..92E, 2016A&A...595A..70E}.
It is seen in Fig.~\ref{fig:sloanscl} that the peak density in the HMC reconstruction is coincident with the Virgo-Coma (Scl\,111) supercluster defined by the concentration of ACO clusters.

%%%%%%%%%%%%%%%%%%%%%%%%%%%%%%%%%%%%%%%
\begin{figure}
    \centering
    \includegraphics[width=1.\linewidth]{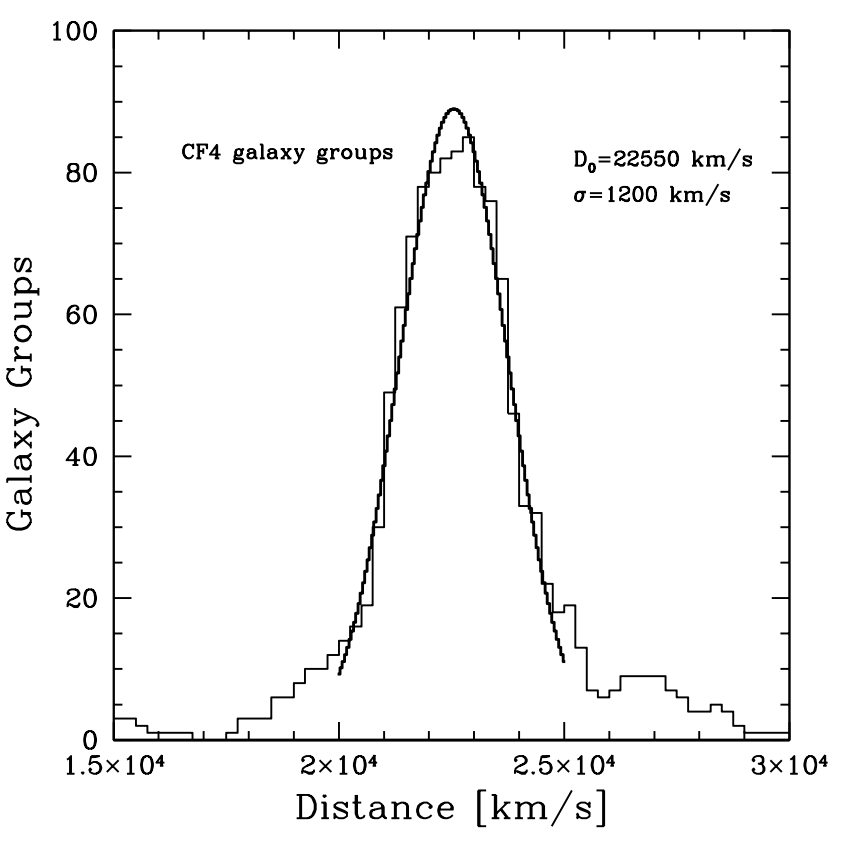}
    \includegraphics[width=1.\linewidth]{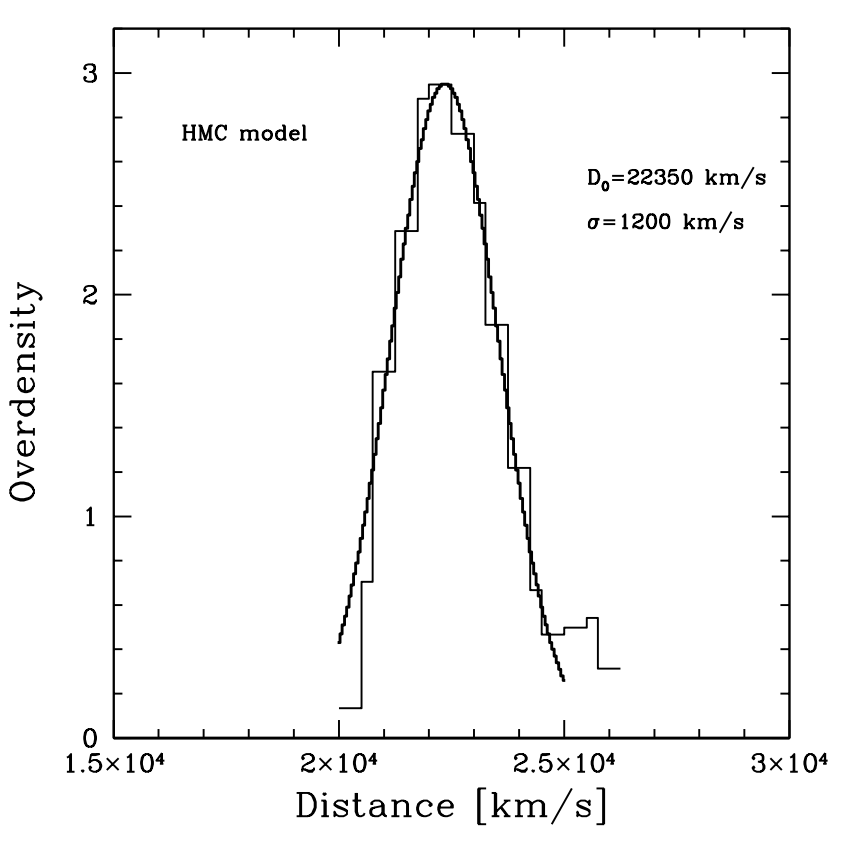}
    \caption{Top: Histogram of CF4 galaxy counts in a beam centered on the Scl\,111 supercluster and a Gaussian fit to the distribution. Bottom: The overdensity from the HMC  model in the same beam with a Gaussian fit.}
    \label{fig:2hist}
\end{figure}
%%%%%%%%%%%%%%%%%%%%%%%%%%%%%%%%%%%%%%%

The run with redshift distance in the number count of CF4 galaxy groups is shown in Fig.~\ref{fig:2hist}.  
In a beam pointing at Abell\,1516 in the Scl\,111 supercluster, galaxy groups in CF4 are counted at 250~\kms\ intervals within 1500~\kms\ spheres (so adjacent steps are not fully independent, resulting in smoothing). 
The distribution at the Sloan Great Wall peak is fitted with a Gaussian centered at 22,550~\kms\ and variance 1,200~\kms. 
The lower panel of the figure shows the run of overdensities in the mean HMC model in the same beam.  The density is taken at the closest cell in the HMC reconstruction at 250~\kms\ intervals along the beam.  The Gaussian fit to the overdensity at the Sloan Great Wall is shifted 200~\kms\ to a smaller distance, with the same variance.

The comparison is not trivial.
The galaxy distribution is in observed redshift space with little error but subject to redshift-space distortion.
The overdensity distribution is inferred from departures from smooth expansion in the velocity field.  
Peculiar velocities infer a density peak that is coincident with the galaxy number counts given a $-200$~\kms\ redshift-space displacement.
The inference of mass from the group counts and the kinematically derived mass from HMC modeling are in accord.

Regarding the properties of component clusters, given the dominance of the Sloan BoA within the CF4 volume, the preponderance of ACO or x-ray clusters is not remarkable. 
Other regions at similar distances have more ACO or x-ray clusters.
Within the study region, there are 28 ACO clusters and 9 x-ray clusters associated with the Shapley supercluster (Scl\,124).  To the south but within the distance of the Sloan BoA, there are 35 ACO clusters and 10 x-ray clusters associated with the Horologium-Reticulum supercluster (Scl\,48) \citep{2001AJ....122.2222E}.

%%%%%%%%%%%%%%%%%%%%%%%%%%%%%%%%%%%%%%%
\begin{figure}
    \centering
    \includegraphics[width=1.\linewidth]{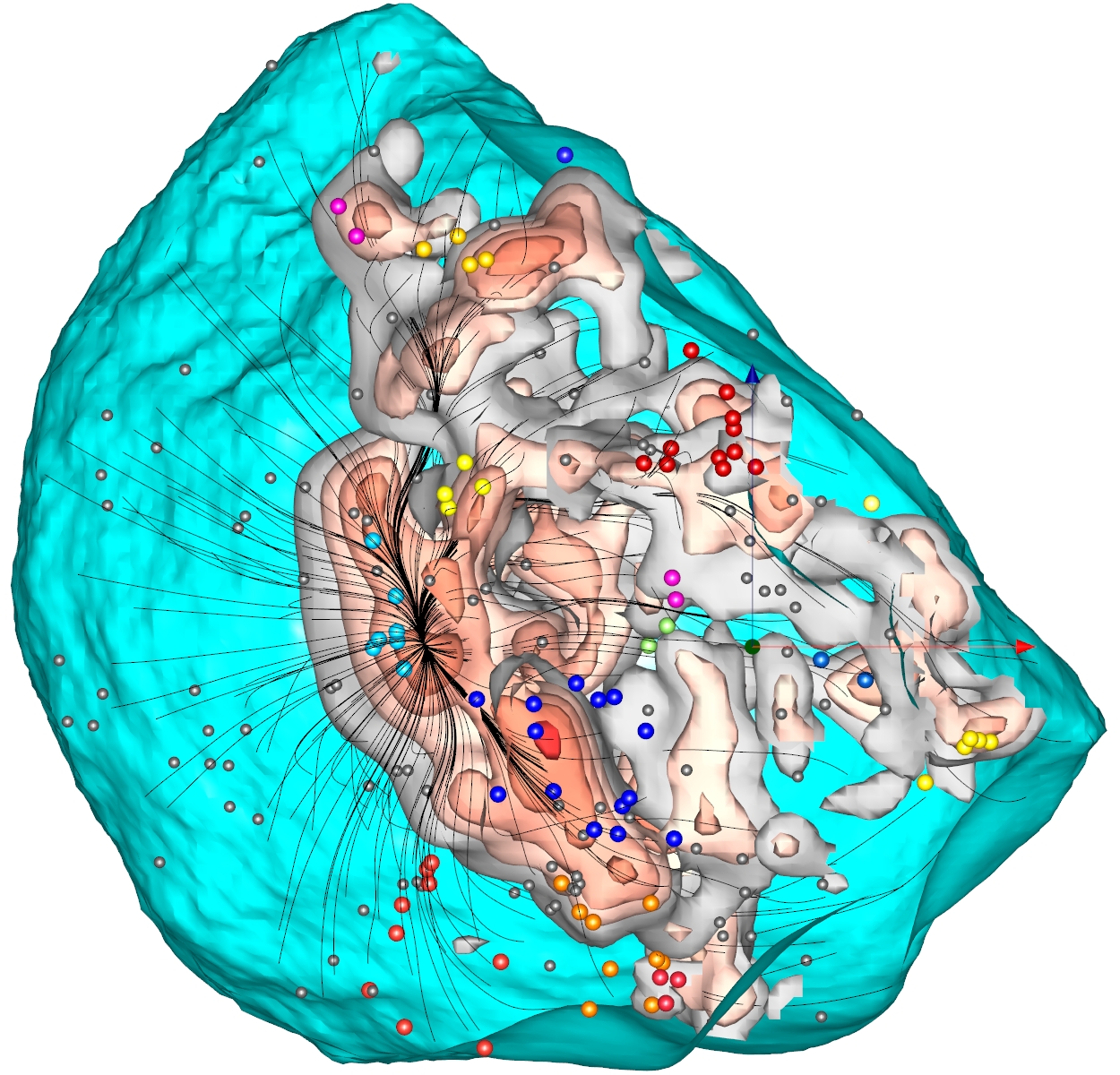}
    \caption{Streamlines seeded throughout the Sloan BoA go to a sink within the Scl\,126 supercluster.
    Scene at 05:00 in video. View it in 3D in this \href{https://sketchfab.com/3d-models/sloan-basin-of-attraction-289e6b9fbd9a44de8c0e959a2a09fc1f}{interactive visualization}.}
    \label{fig:sloanstreamlines}
\end{figure}
%%%%%%%%%%%%%%%%%%%%%%%%%%%%%%%%%%%%%%%

Another revelation comes from an inspection of streamlines sourced throughout the Sloan BoA, seen at 04:58 in the video and Fig.~\ref{fig:sloanstreamlines}.
The streamlines terminate not at SCL\,111, the site of highest reconstructed density and largest accumulation of ACO clusters, but rather at SCL\,126 with only half as many ACO clusters and no x-ray clusters.

%%%%%%%%%%%%%%%%%%%%%%%%%%%%%%%%%%%%%%%
\begin{figure}
    \centering
    \includegraphics[width=1.\linewidth]{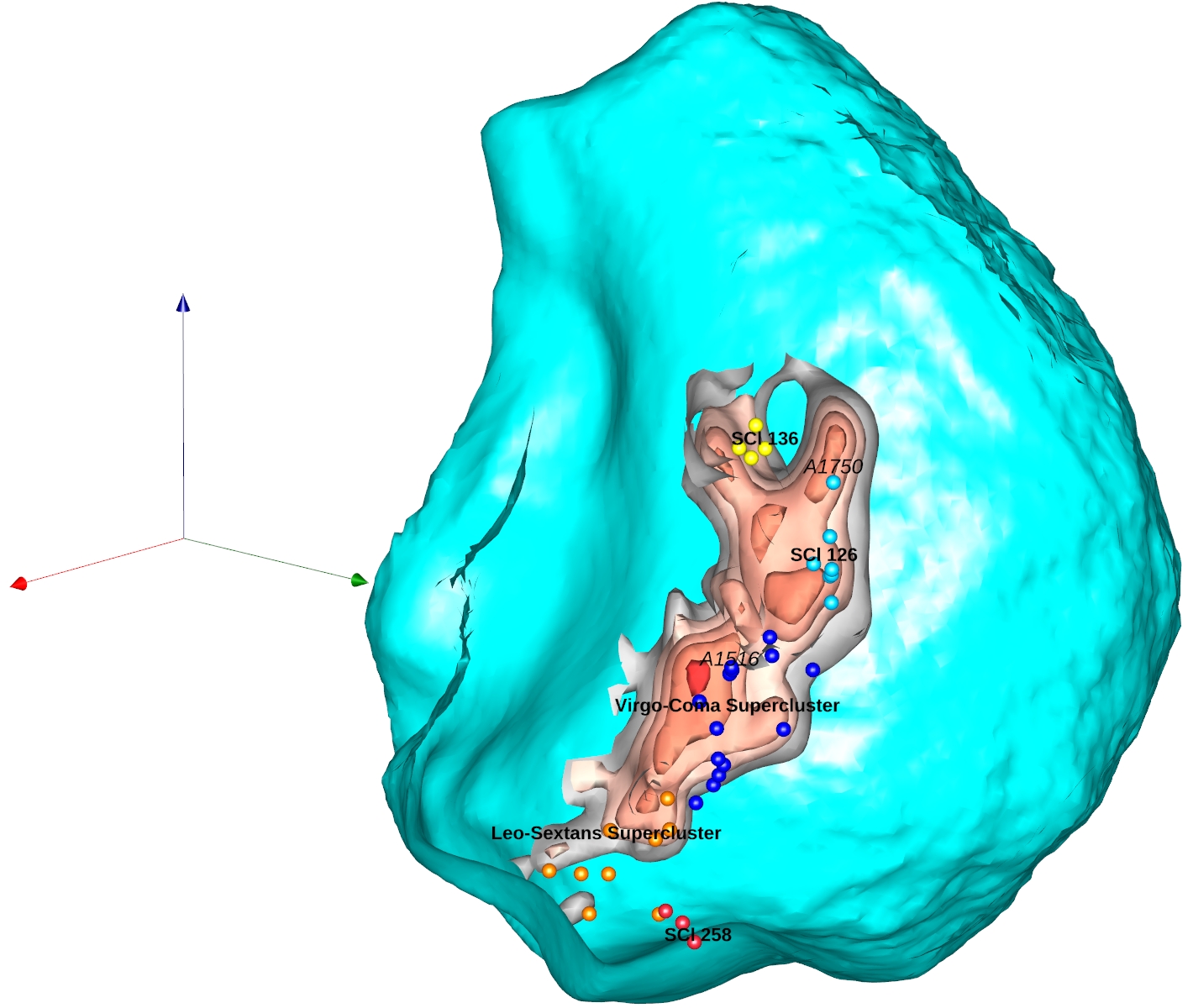}
    \caption{Density contours isolated to the spine of the Sloan Great Wall.  ACO clusters are located with colored balls (Scl\,136: yellow; Scl\,126: cyan; Scl\,111: blue; Scl\,91: tan; Scl\,258: red). Scene at 06:10 in video. 
    %View it in 3D in this \href{https://sketchfab.com/3d-models/sloan-basin-of-attraction-the-spine-27045a79a65b4c10b11e31f22a8d13e9}{interactive visualization}.
    }
    \label{fig:spine}
\end{figure}
%%%%%%%%%%%%%%%%%%%%%%%%%%%%%%%%%%%%%%%

The spine of the Sloan Great Wall, shown in Fig.~\ref{fig:spine}, runs through Scl\,136, Scl\,126, Virgo-Coma=Scl\,111, Leo-Sextans=Scl\,91, and Scl\,258.
\citet{2011ApJ...736...51E, 2016A&A...595A..70E} have discussed the global morphologies and galactic properties of the main Sloan Great Wall components Scl\,111 and Scl\,126 in considerable detail.
They describe Scl\,126 as a rich filament and Scl\,111 as a multi-spider, with high density regions connected by filaments.
There are differences in the relative populations of spirals and ellipticals between supercluster regions with the fraction of red galaxies largest in the core of the supercluster Scl 126.
They suggest that larger peculiar velocities of galaxies in Scl\,126 are indicative of more active merging.
Scl\,126 may be less dynamically evolved than Scl\,111.
It can be speculated if this apparent dynamical activity associated with Scl\,126 is related to its situation as the evident sink of Sloan BoA streamlines.

%%%%%%%%%%%%%%%%%%%%%%%%%%%%%%%%%%%%%%%
\begin{figure}
    \centering
    \includegraphics[width=1.\linewidth]{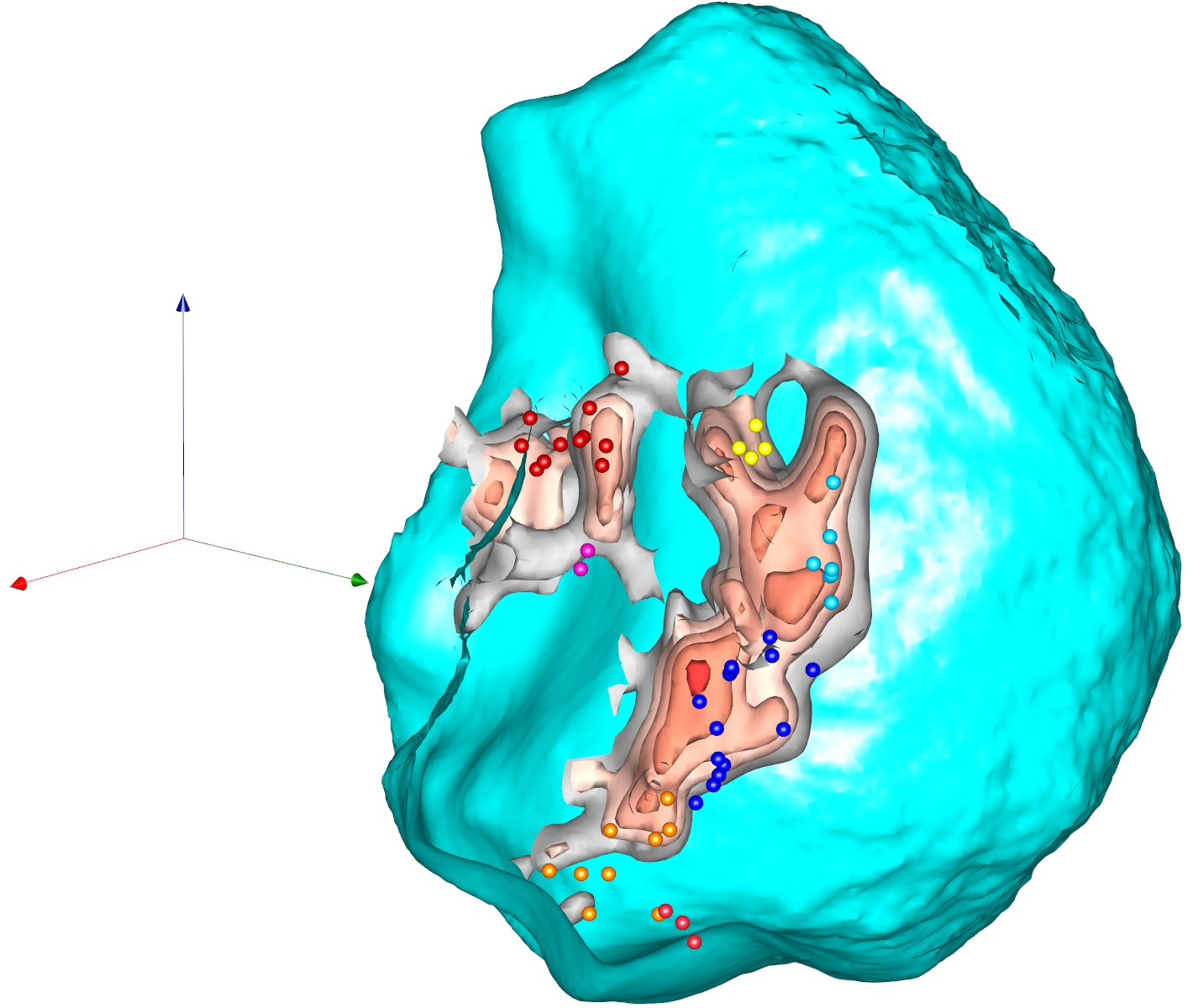}
    \caption{Density contours are extended to include the Bo\"otes supercluster structure.  Associated ACO clusters are shown as red balls; purple balls are within the adjacent Scl\,122. Scene at 06:14 in video. View it in 3D in this \href{https://sketchfab.com/3d-models/sloan-basin-of-attraction-the-spine-and-bootes-9bdfed73b996497183a5b43d621fc7eb}{interactive visualization}.}
    \label{fig:bootes}
\end{figure}
%%%%%%%%%%%%%%%%%%%%%%%%%%%%%%%%%%%%%%%

The Bo\"otes supercluster, Scl\,138, is a secondary over dense structure within the Sloan BoA at $p=0.5$ (seen in Fig~\ref{fig:bootes} as the structure added to that in Fig.~\ref{fig:spine}). With 12 ACO clusters, it is the location of the second greatest accumulation of rich clusters in the Sloan BoA.
X-ray observations of the dominant Abell\,1795 indicate this cluster is reasonably relaxed \citep{2023A&A...678A..91K}.  
The nearby Abell\,1775 is interesting: it contains an 800~kpc head-tail continuum radio structure and an extended fossil x-ray plasma plausibly due to a cluster merging event \citep{2021A&A...649A..37B}. 

The discussion will return to the Bo\"otes supercluster in connection with the later section regarding the Ho`oleilana baryon acoustic oscillation structure.

%%%%%%%%%%%%%%%%%%%%%%%%%%%%%%%%%
\section{Other Basins of Attraction}

It was seen in Fig.~\ref{fig:p0.5} that there are four BoA adjacent to the Sloan BoA: identified as Corona Borealis+19.6 in blue, Serpens Caput+26.2 in yellow, Hercules+28.6 in red, and Leo+23.2 in orange.  
These entities are shown at $p=0.5$ and are quite stable with respect to variations in probability level.   
The only other substantial structures to be given attention are Bo\"otes+21.6 (distinct from the Bo\"otes supercluster discussed above) and Corona Borealis+23.7 (also known in the literature as Corona Borealis supercluster).

%%%%%%%%%%%%%%%%%%%%%%%%%%%%%%%%%
\subsection{Serpens Caput+26.2 and Hercules+28.6 Basins of Attraction}

\begin{figure}
    \centering
\includegraphics[width=1.\linewidth]{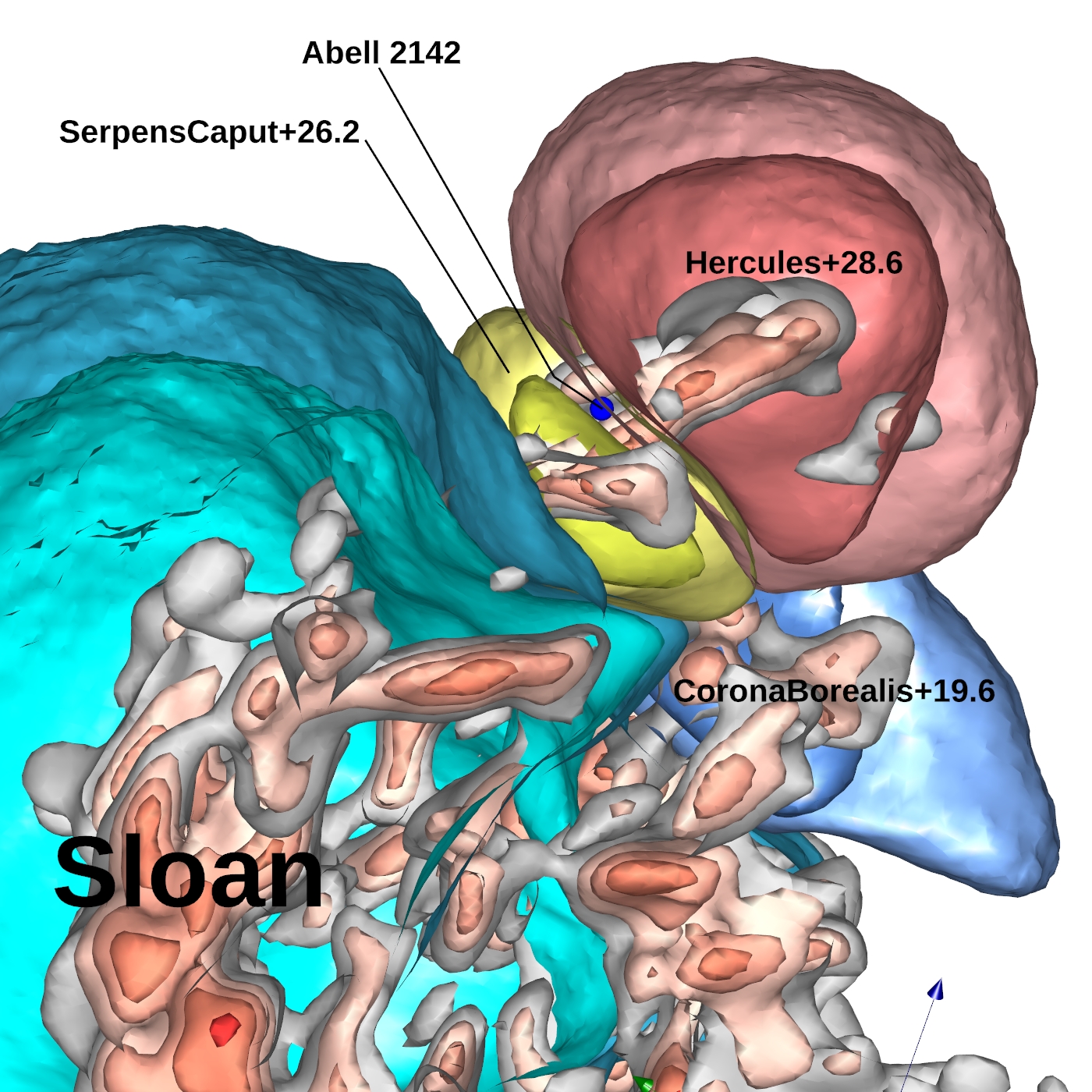}
    \caption{Serpens Caput+26.2 and Hercules+28.6 BoA are shown with $p=0.5$ and 0.3 isosurfaces in dark and light shades of yellow and red respectively. The rich cluster Abell\,2142 is located in the Serpens Caput+26.2 BoA very close to Hercules+28.6. The Sloan Great Wall BoA lies immediately to the lower left, with the same $p$ levels in cyan and blue.
    Scene at 07:06 in video.
    View it in 3D in this \href{https://sketchfab.com/3d-models/sloan-basin-of-attraction-and-neighborhood-65fcd0eafed94377a6b9c4bc5fbb71b5}{interactive visualization}.}
    \label{fig:cercap}
\end{figure}

Two adjacent BoA are reasonably stable
to variations in the probability $p$ level, seen respectively with yellow and red $p=0.5$ isosurfaces in Fig.~\ref{fig:p0.5}. 
The region is isolated in Fig.~\ref{fig:cercap}. It is seen that the BoA domains increase only modestly in expanding from $p=0.5$ to $p=0.3$, abutting without impinging on each other. Similarly, Serpens Caput+26.2 and Sloan Great Wall abut without impinging.\footnote{By definition $p=0.5$ BoA cannot overlap. Locations with $p<0.5$ can potentially lie in alternative BoA.}

The cluster Abell\,2142 and associated supercluster Scl\,A2142 has received considerable attention \citep{2015A&A...580A..69E, 2018A&A...620A.149E, 2020A&A...641A.172E}. It lies in the Serpens Caput+26.2 BoA but near the boundary with Hercules+28.6.   
The most notable feature is a straight filament of projected length 3,600~\kms\ emanating from A2142.
The series of three papers discuss the "cocoon" of low density isolating the supercluster and the morphologies of component galaxies, remarking on the distributions of old and young populations and of active nuclei.

%%%%%%%%%%%%%%%%%%%%%%%%%%%%%%%%%%%%%%%
%%%%%%%%%%%%%%%%%%%%%%%%%%%%%%%%%%%%%%%
%%%%%%%%%%%%%%%%%%%%%%%%%%%%%%%%%
\subsection{Corona Borealis+19.6 Basin of Attraction and Corona Borealis+23.7 Possible Basin of Attraction}

Close in the line-of-sight and to the foreground of the two BoA just discussed are Corona Borealis+19.6 BoA and the prospective Corona Borealis+23.7 BoA.  This latter entity appears at $p=0.34$.  
Within Corona Borealis+23.7 lies the well studied Corona Borealis supercluster \citep{1988AJ.....95..267P, 1998ApJ...492...45S, 2014MNRAS.441.1601P, 2021A&A...649A..51E}.
There are 8 ACO clusters and 2 x-ray clusters associated with the core Scl\,158, including the dominant Abell\,2065 and adjacent A2061, A2067 and A2089.
An important cluster that did not make the ACO catalog is GR2064 \citep{2021A&A...649A..51E}. 

The individual galaxy and large scale structural morphologies, and inferred dynamics were extensively studied by \citet{2021A&A...649A..51E}. 
Very old stellar populations are found at the centers of all the rich clusters.
Filaments connect between the central A2065 and each of A2089 and A2061 and in turn between A2061 and A2067.
Mixes of stellar populations, galaxies that can be associated with blue star formation, red quiescent, and green valley intermediate stages, are found in the filaments and field around the clusters.
Some younger populations are found within the clusters, attributable to recent arrivals.
Of particular interest, the domains of infall for each of the major clusters could be isolated from  the runs of galaxy density with radius around the clusters.
Most likely, A2067 with $10^{14} \Msun$ will merge with $10^{15} \Msun$ A2061 at $3h^{-1}$~Mpc in projection in a few Gyr.
To paraphrase \citet{2021A&A...649A..51E}, in a Hubble time, all four of the Abell clusters in the Corona Borealis supercluster may merge to form one of the largest bound structures in the nearby Universe.

%%%%%%%%%%%%%%%%%%%%%%%%%%%%%%%%%%%%%%%
\begin{figure}
    \centering
\includegraphics[width=1.\linewidth]{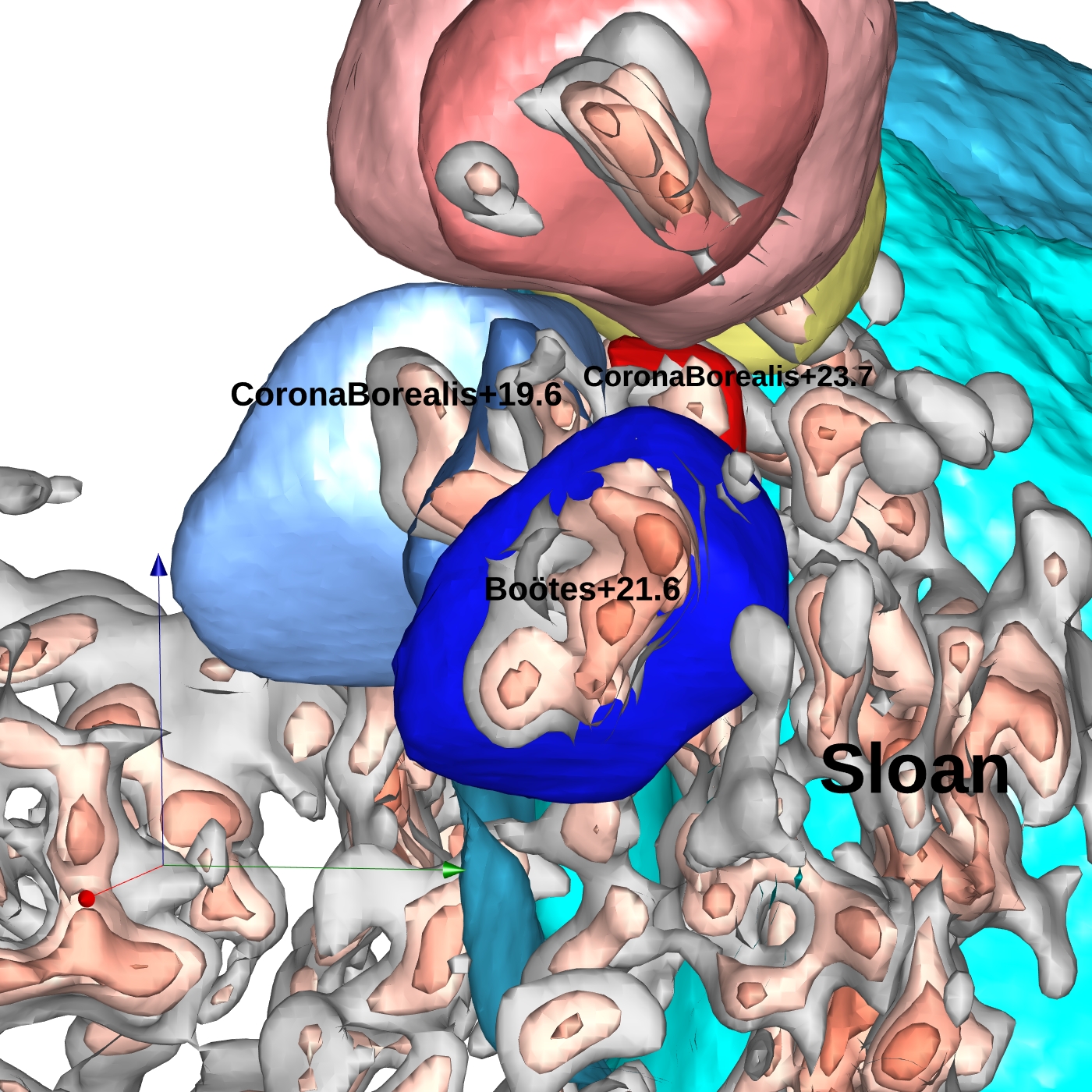}
    \caption{The Corona Borealis+19.6 with $p=0.5$ and 0.3 levels in dust blue and grey and Corona Borealis+23.7  with the $p=0.3$ level in red are behind the prospective Boötes+21.6 BoA at $p=0.3$ in violet. Sloan Great Wall BoA surfaces in cyan and blue are at lower right.
    Scene at 07:15 in video
    View it in 3D in this \href{https://sketchfab.com/3d-models/sloan-basin-of-attraction-and-neighborhood-65fcd0eafed94377a6b9c4bc5fbb71b5}{interactive visualization}.}
    \label{fig:corbor}
\end{figure}
%%%%%%%%%%%%%%%%%%%%%%%%%%%%%%%%%%%%%%

Returning to Corona Borealis+19.6, it is seen in Fig.~\ref{fig:corbor} that the BoA isosurface at $p=0.3$ extends to smaller velocities than the 14,000~\kms\ edge of the study area. This edge rises almost vertically left of center in the figure, with the dust blue BoA isosurface extending to the left).
Here, the BoA is expanding at low $p$ to include the Bo\"otes Void. 

%%%%%%%%%%%%%%%%%%%%%%%%%%%%%%%%%%%%%%
\subsection{Bo\"otes+21.6 Possible Basin of Attraction}

Bo\"otes+21.6 is realized as a BoA at $p=0.46$.
It is centered on Scl\,138 (Bo\"otes) supercluster with 12 Abell clusters.
At lower probability it can extend to Scl\,109 (Ursa Major) supercluster with 8 Abell clusters.
There is considerable ambiguity in the region involving Corona Borealis+19.6, Corona Borealis+23.7, Bo\"otes+21.6, running down to the Ursa Major supercluster.
It will be seen in the next section that networks of high density filaments connect all these features.
With different probabilities, these structures can stand as separate BoA or bleed into one another or into the Sloan Great Wall BoA.

%%%%%%%%%%%%%%%%%%%%%%%%%%%%%%
\subsection{Leo+23.2 Basin of Attraction}

The Leo+23.2 BoA is seen in Fig.~\ref{fig:p0.5} to lie on an edge of the Sloan Great Wall BoA well removed from the other structures that have been discussed.

%%%%%%%%%%%%%%%%%%%%%%%%%%%%%%%%
\section{V-Web Structures}

%%%%%%%%%%%%%%%%%%%%%%%%%%%%%%%%%%%%%%%
\begin{figure}
    \centering
\includegraphics[width=1.\linewidth]{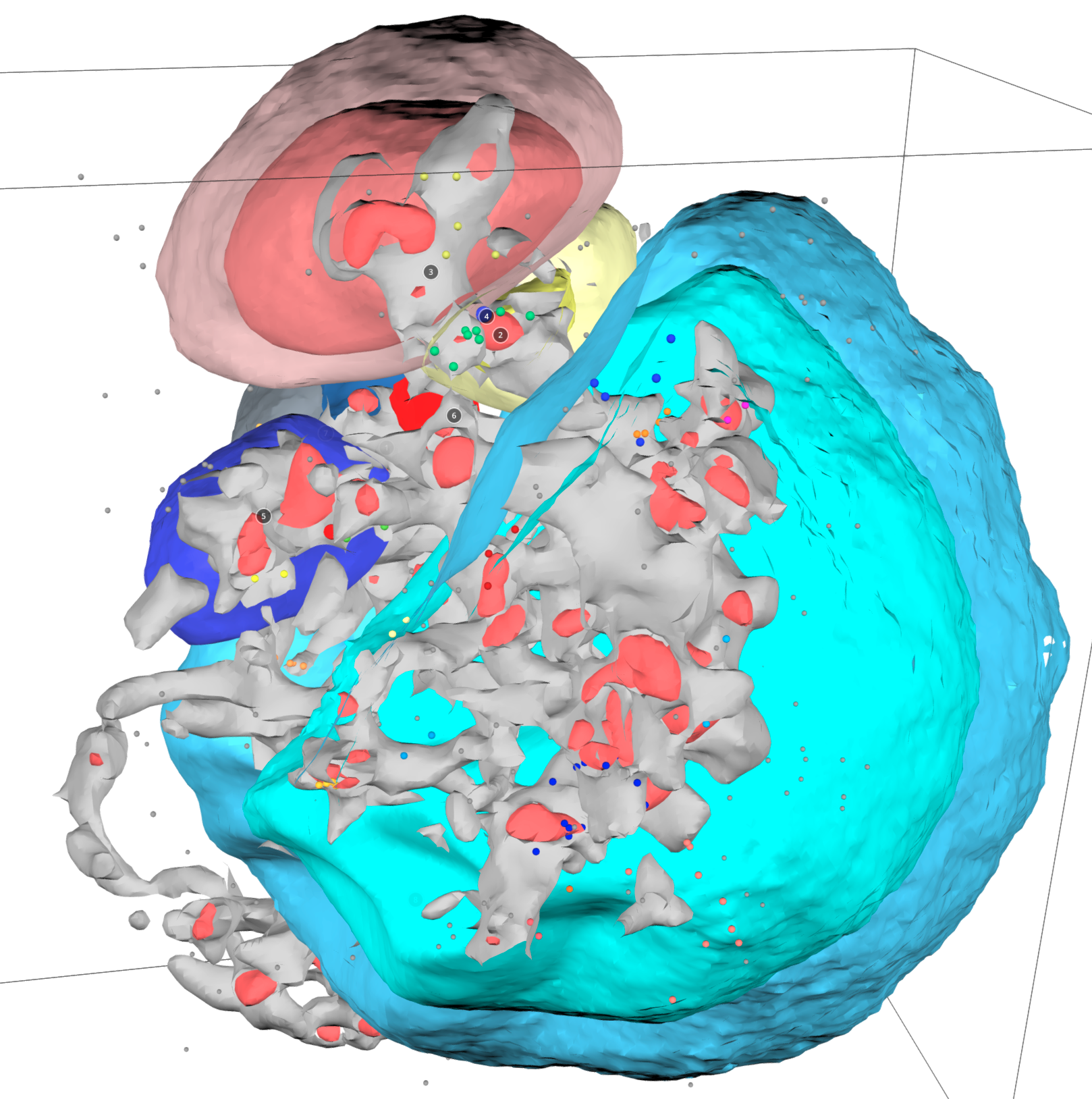}
    \caption{Knots (red) and filaments (gray) defined by the eigenvalues of the velocity shear provide an alternative description of the density field.  The V-web filaments and knots are shown within BoA isosurfaces at $p=0.5$ and 0.3 levels. View it in 3D in this \href{https://sketchfab.com/3d-models/sloan-basin-of-attraction-neighborhood-v-web-355ed906610647c59d2ff9bb0a5b693a}{interactive visualization}. Scene at 08:12 in video.}
    \label{fig:vweb}
\end{figure}
%%%%%%%%%%%%%%%%%%%%%%%%%%%%%%%%%%%%%%

%%%%%%%%%%%%%%%%%%%%%%%%%%%%%%%%%%%%%%%
\begin{figure}
    \centering
\includegraphics[width=1.\linewidth]{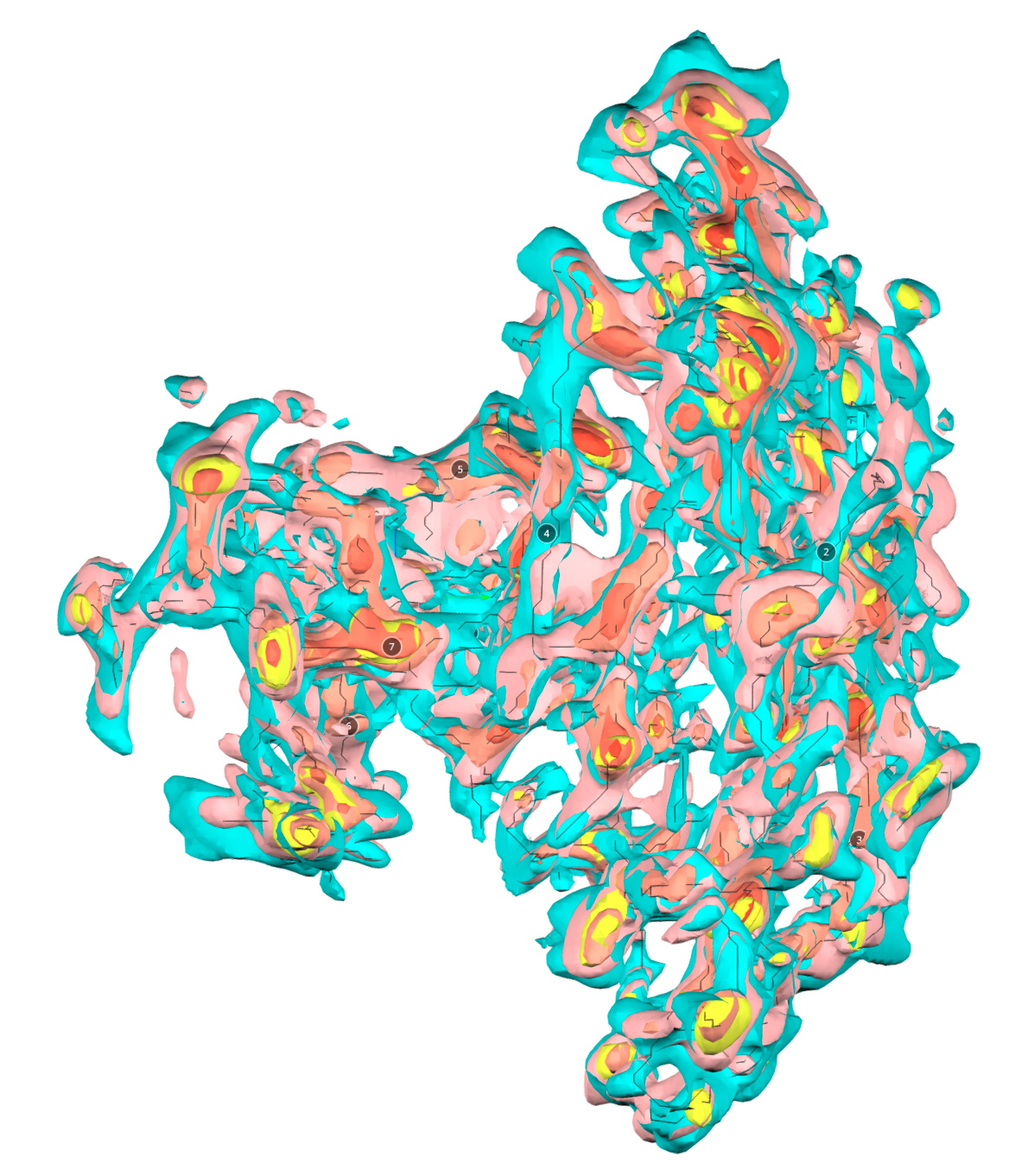}
    \caption{A superposition of V-web filaments and knots on the HMC density contours for the full CF4 volume.  High density HMC levels are shown as pink and red isosurfaces and V-web filaments and knots are shown in cyan and yellow, respectively. View it in 3D in this \href{https://sketchfab.com/3d-models/cf4-v-web-filament-finder-density-isos-f164e9d202744bb0a260600d1c9322dc}{interactive visualization}. Scene at 08:55 in video.}
    \label{fig:vweb_HMC}
\end{figure}
%%%%%%%%%%%%%%%%%%%%%%%%%%%%%%%%%%%%%%

Large scale structure can be characterized by shear in the velocity field.  The theory underlying the so-called velocity or V-web 
 was presented by \citet{2012MNRAS.425.2049H} and applied to reconstruction of the LSS by \citet{2014MNRAS.441.1974L}. \citet{2017ApJ...845...55P} provided an application with a map of observed peculiar velocities inferred from the Cosmicflows-2 Catalog, as well as with the Cosmicflows-3 Catalog in the context of the discovery of the South Pole Wall \citep{2020ApJ...897..133P}.
Given the  mean velocity field following from the HMC analysis the shear tensor can be calculated:

\begin{equation}
    \Sigma_{\alpha\beta} = -(\partial_{\alpha}V_{\beta} + \partial_{\beta}V_{\alpha})/2H_0
    \label{eq:shear}
\end{equation}
where partial derivatives of the velocity $V$ are determined along the orthogonal directions $\alpha$ and $\beta$.  $H_0$ provides a normalization of the average universal expansion.  The negative sign associates positive eigenvalues with collapse.
The three orthogonal eigenvalues describe whether a location is in a knot in overall collapse, a filament collapsing on two axes, a sheet collapsing on only one axis, or a void in overall expansion.

Knots and filaments delineate the regions of highest density. Setting cells with the eigenvalues of sheets and voids to the background reveals networks of filaments and knots.  The geometric centers of filaments can be located by the COsmic Web Skeleton (COWS) method described by 
\citet{2022MNRAS.514..470P}.
Adjacent centers can be connected into the skeleton of a filament.

The interconnectedness of filaments is found with remarkable clarity in the V-web analysis.
As seen in Fig.~\ref{fig:vweb} and the accompanying interactive model, there are robust connections between every BoA and at least one of its neighbors.  In such cases, there are reversals of streamlines at the junctions between BoA.
This phenomenon of flow reversals within filaments at the boundaries separating BoA is well documented; for example see Fig.~5 in \citet{2017ApJ...845...55P} and causes discontinuities in the COWS representation.

The V-web and HMC high density structures are shown in superposition in Fig.~\ref{fig:vweb_HMC}.
Most notably, the HMC structures are contained within separate BoA whereas the V-web filaments connect across BoA boundaries creating a universal network.

%%%%%%%%%%%%%%%%%%%%%%%%%%%%%%%%%
\section{Ho`oleilana Baryon Acoustic Oscillation Structure}

\begin{figure}
    \centering
\includegraphics[width=1.\linewidth]{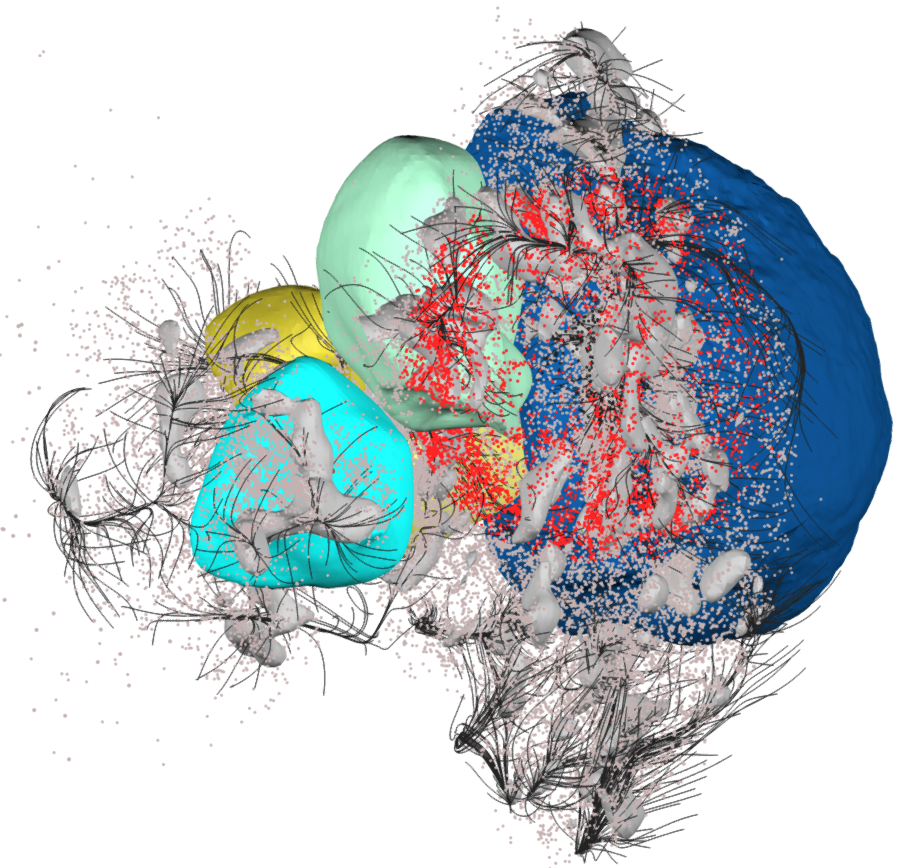}
    \caption{The Ho`oleilana Baryon Acoustic Oscillation feature is superimposed on the Hamiltonian Monte Carlo Basin of Attraction structures. The primary known BoA surfaces are shown at p=0.5: SGW BoA (blue), Hercules-CfAGW (Green), Perseus-Pisces (cyan), Shapley (yellow), and Ophiuchus (gold). Black velocity streamlines seeded randomly accumulate at BoA cores. Galaxies from the CF4 sample are represented by small circles; red if associated with the Ho`oleilana BAO shell, black if associated with the Boötes supercluster core, and gray otherwise. 
    View it in 3D in this \href{https://sketchfab.com/3d-models/cf4-p-boas-hooleilana-179fb312fffe417c8b6e750c8cff3522}{interactive visualization}. Scene at 10:46 in video.}
    \label{fig:hooleilana}
\end{figure}

The Sloan Great Wall BoA is large but the regions of over density are only a component of the Ho`oleilana baryon acoustic oscillation feature \citep{2023ApJ...954..169T}.  Indeed, as shown in Fig.~\ref{fig:hooleilana}, the Ho`oleilana feature extends to the foreground to pass through the Hercules and CfA Great Wall structures.  The Sloan Great Void is within its embrace.

%%%%%%%%%%%%%%%%%%%%%%%%%%%%%%%%%%%%%%%
%%%%%%%%%%%%%%%%%%%%%%%%%%%%%%%%%%%%%%

It was argued by \citet{2023ApJ...954..169T} that the probability that Ho`oleilana is a real BAO shell is greater than 99\%.  Accordingly, the incipient dark matter core is to be associated with the Bo\"otes supercluster.
However, HMC velocity streamlines do not terminate at this supercluster but, rather, gather within the Sloan Great Wall at the Scl\,126 supercluster.

The shell of Ho`oleilana passes across BoA boundaries as freely as V-web filaments.
Such must be a general property of BAO given their $\sim 100/h$~Mpc radius scales, much larger than observed BoA scales.
Today's residues of baryon acoustic oscillation structures are individually fragile, the consequences of tidal mixing on grand scales. 

%%%%%%%%%%%%%%%%%%%%%%%%%%%%%%%%
\section{Discussion and Summary}
%%%%%%%%%%%%%%%%%%%%%%%%%%%%%%%%

This article complements the presentation by \citet{2024NatAs.tmp..234V} of the interpretation of Cosmicflows-4 (CF4) distances and inferred peculiar velocities with a Hamiltonian Monte Carlo reconstruction of density and velocity fields.
Flow patterns begun at arbitrary locations (sources) can be traced along streamlines to sinks.
A volume that encompasses all streamlines terminating in a common sink is a basin of attraction (BoA).
If the distances and velocities of galaxies were known without error and with sufficient density, the boundaries of BoA would be uniquely defined and the Universe could be segmented into abutting BoA each with the mean density.
However, measurement uncertainties are substantial, with the consequence that boundaries are only established in probablistic terms. 

The paper by \citet{2024NatAs.tmp..234V} gave particular attention to the domain of CF4 coverage within $z=0.05$ where there is reasonable all-sky coverage outside the zone of obscuration. 
The most prominent BoA in that inner region are Shapley, Hercules, Perseus-Pisces, and Ophiuchus.
Most probably, the Milky Way is linked to the Shapley BoA although there is a non-negligible possibility of linkage to Ophiuchus BoA.

Here, the discussion turns to the domain with $z>0.05$, uniquely covered in CF4 in the north Galactic, north celestial sector of the sky drawing primarily on the Sloan PV catalog of \citet{2022MNRAS.515..953H} with FP distances.
The separation from the previous study benefits from the gap in over dense structures with voids at 11,000-14,000~\kms.
Indeed, there is a large contiguous void that parallels the Sloan Great Wall that we call the Sloan Great Void.
The famous Bo\"otes Void \citep{1987ApJ...314..493K} is an appendage.

Five BoA are identified in the study region but the largest by far is the Sloan BoA.
It may well extend beyond the CF4 data domain to the celestial south.
The discussion focuses on BoA domains at probability $p=0.5$ but the domains in the study sector are reasonably stable with relaxation of the probability $p$ levels.

The gravitating mass, dark plus baryonic, within the Sloan BoA can be crudely estimated in two ways that give similar results.
By $p=0.3$ the BoA in the study region are filling most of space without significant overlaps.
One estimate assigns mass to cells within the BoA outline, $(1+\delta) M_0$, where $\delta$ is the over-density in a cell from the HMC model and $M_0$ is the mass in a cell at the mean density of the Universe,  then the sum is taken over all the cells in the volume.
An alternate estimate is to simple take the product of the mean universal density in gravitating matter times the BoA volume following from the definition of a BoA and assuming the observed volume is a good representation.
The two methods both give a mass of $3\times 10^{18}~\Msun$.
This is an impressive $10^{-5}$ of the observable Universe.

There is good agreement between the observed distribution of galaxies in redshift space and inferred HMC density structures although there are anomalies.
Streamlines in the Sloan BoA terminate, not at the highest density peak in the Virgo-Coma Scl\,111 supercluster but in the secondary SCl\,126 supercluster.
Prominent supercluster structures that have received attention in publications  such as Corona Borealis \citep{2021A&A...649A..51E} and Scl\,A2142 \citep{2020A&A...641A.172E} lie in peripheral structures in the present analysis.
The concentration of rich (Abell) and x-ray clusters to the Sloan BoA is not remarkable.
Abundances are greater in other regions with similar redshifts.

The V-web \citep{2012MNRAS.425.2049H} provides an alternative description of large scale structure. The analysis is based on the CF4 catalog of galaxy distances and velocities.
The HMC approach builds probabilistic global density and velocity fields compatible in $\Lambda$CDM cosmology with the ensemble of observed galaxy positions and line-of-sight velocities.
The V-web is constructed from local gradients in the mean velocity field.
HMC high density structures and V-web filaments overlay.
A major difference is that HMC structures are confined within separate BoA while the V-web filaments connect through adjacent BoA.
There are flow reversals within a filament at BoA boundaries, manifesting in breaks in the COWS spine along the filament.

The filament network is complex and difficult to disentangle.
The videos in the Appendix explore five different contiguous paths from the Milky Way to the Sloan BoA.
Many other paths could be found.
The interactive model shows less confusing skeletons of the five paths.

A final section touches on Ho`oleilana \citep{2023ApJ...954..169T}, probably a remnant baryon acoustic oscillation and surely the closest such structure that will be found.
As with the V-web, Ho`oleilana extends across BoA boundaries.
The presumptive dark matter source of the BoA is to be associated with the Bo\"otes supercluster.
However, at probability $p=0.5$, velocity streamlines sourced in the Bo\"otes supercluster flow to a sink at supercluster Scl\,126 in the Sloan Great Wall

There is the caveat that the domains isolated as BoA depend on the scale of the Gaussian smoothing of the velocity field as demonstrated by \citet{2023A&A...678A.176D} and \citet{2026OJAp....957824S}.  Smoothing on smaller scales can cause fragmentation.  As a specific example, streamlines sourced at the Milky Way lead to sinks at the Virgo Cluster with smoothing $<2 h^{-1}$~Mpc, within Laniakea with smoothing $2 h^{-1}$ to $4 h^{-1}$~Mpc, and to the Shapley concentration with smoothing $>4 h^{-1}$~Mpc with the \citet{2026OJAp....957824S} analysis.
With the \citet{2020MNRAS.493.3513D} analysis the BoA properties stabilize at smoothing $>10 h^{-1}$~Mpc.

The BoA on the largest scales are of particular cosmological interest.  For example, are they consistent with $\Lambda$CDM theory?  It is expected that the streamline sinks of these largest scale BoA would be at rest in the CMB frame.  Manifestly, the Virgo Cluster sink is not at rest in the CMB frame and, most likely, neither is a Laniakea sink at rest in the CMB frame.  A Shapley (or Sloan) sink might be, but current data constraints are insufficient.  

The present HMC study \citep{2024NatAs.tmp..234V} is based on linear theory with a minimum resolution of $5 h^{-1}$~Mpc.  To fully understand the implications of the smoothing scale in this work, the same methodology should be applied to Gpc-scale simulations, with and without distance and velocity uncertainties.  Pressing on a longer term is the need for quasi-all-sky galaxy distance/peculiar velocity surveys that explore volumes much larger than CF4 to empirically establish the sizes of the largest BoA in the local Universe.

\bigskip
\begin{acknowledgements}
This paper would not have been possible without the contributions of our colleagues in the assembly of the Cosmicflows-4 data and in the subsequent analysis: H\'el\`ene Courtois, Ehsan Kourkchi, Cullan Howlett, Khaled Said, Simon Pfeifer, and many others.
YH is partially supported by the Israel Science Foundation (ISF 1450/24).
\end{acknowledgements}

%%%%%%%%%%%%%%%%%%%%%%%%%%%%%%%
\bibliography{biblio}{}
\bibliographystyle{aasjournal}
%%%%%%%%%%%%%%%%%%%%%%%%%%%%%%%%
\appendix
%%%%%%%%%%%%%%%%%%%%%%%%%%%%%%%%

\begin{figure}
    \centering
\includegraphics[width=\linewidth]{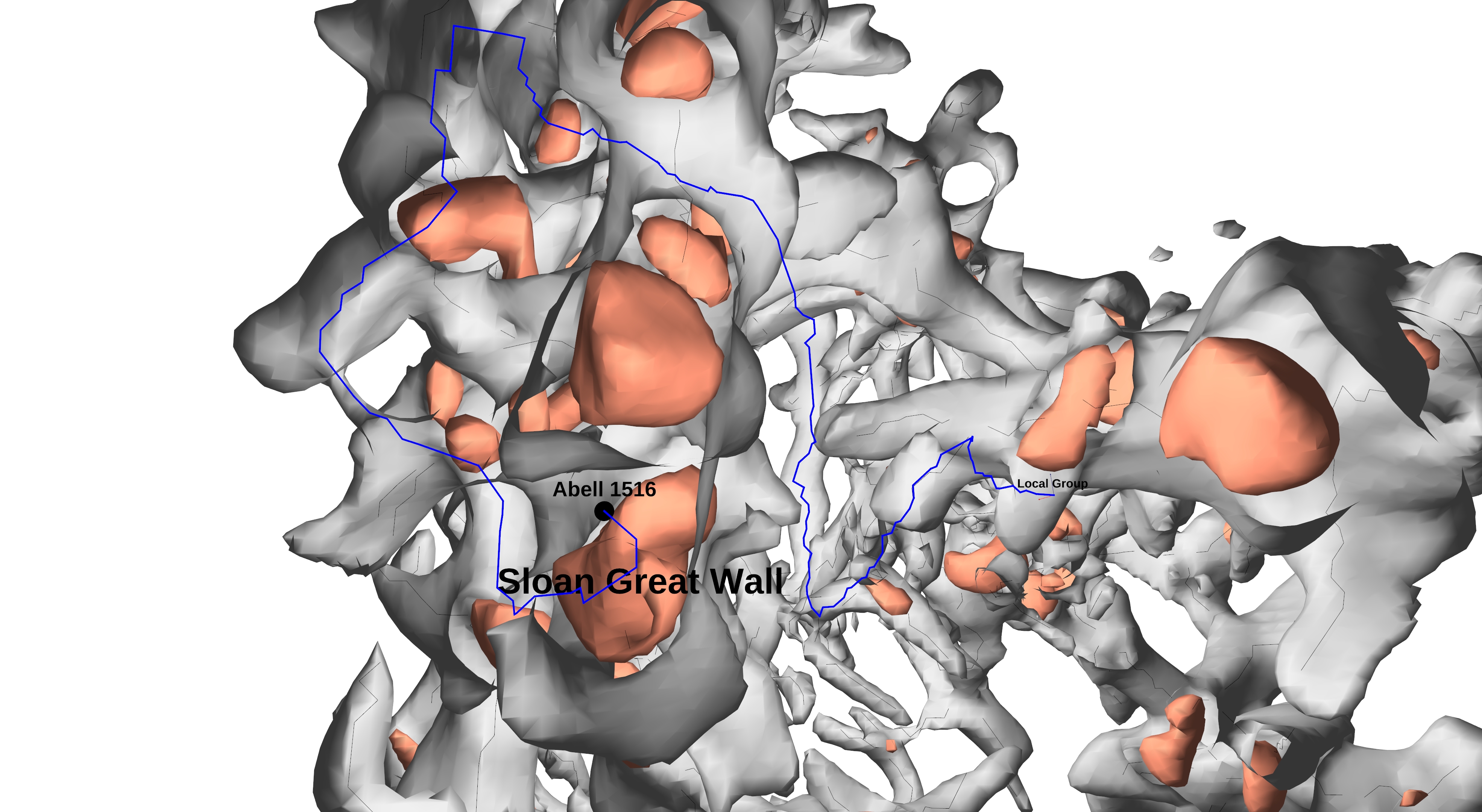}
    \caption{Daniel's Path (blue) along the V-web from the Milky Way to the Sloan BoA. This path is explored in this \href{https://vimeo.com/1068786209/e97b1b6c3c}{video}. Notable features along this path, the northernmost path of all paths, are the CfA Great Wall, Coma cluster, Dragon's Tail filament, Corona Borealis Supercluster, and Boötes+30 Wall.}
    \label{fig:path_daniel}
\end{figure}

\begin{figure}
    \centering
\includegraphics[width=\linewidth]{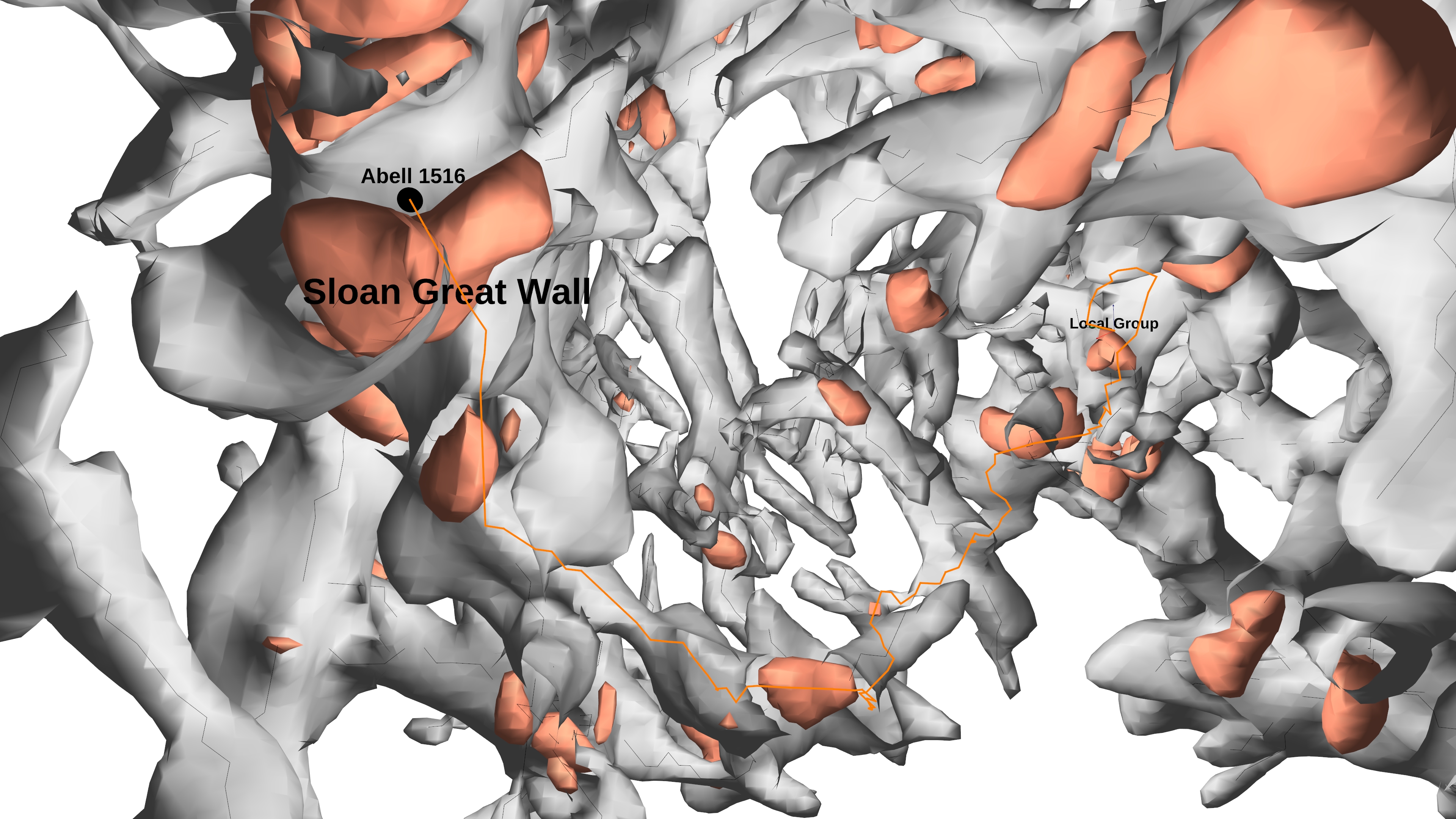}
    \caption{Brent's Path (orange) along the V-web from the Milky Way to the Sloan BoA is explored in this \href{https://vimeo.com/1068800633/4ee79d5545}{video}.  After a short detour through the galactic south across the Cen-Pup-PP filament and Perseus cluster, the path follows a complex network of filaments crossing the Zone of Avoidance down to an X-chromosome shaped knot, a branch of which leads to the Sextans entry point to Sloan Basin of Attraction.}
    \label{fig:path_brent}
\end{figure}

\begin{figure}
    \centering
\includegraphics[width=\linewidth]{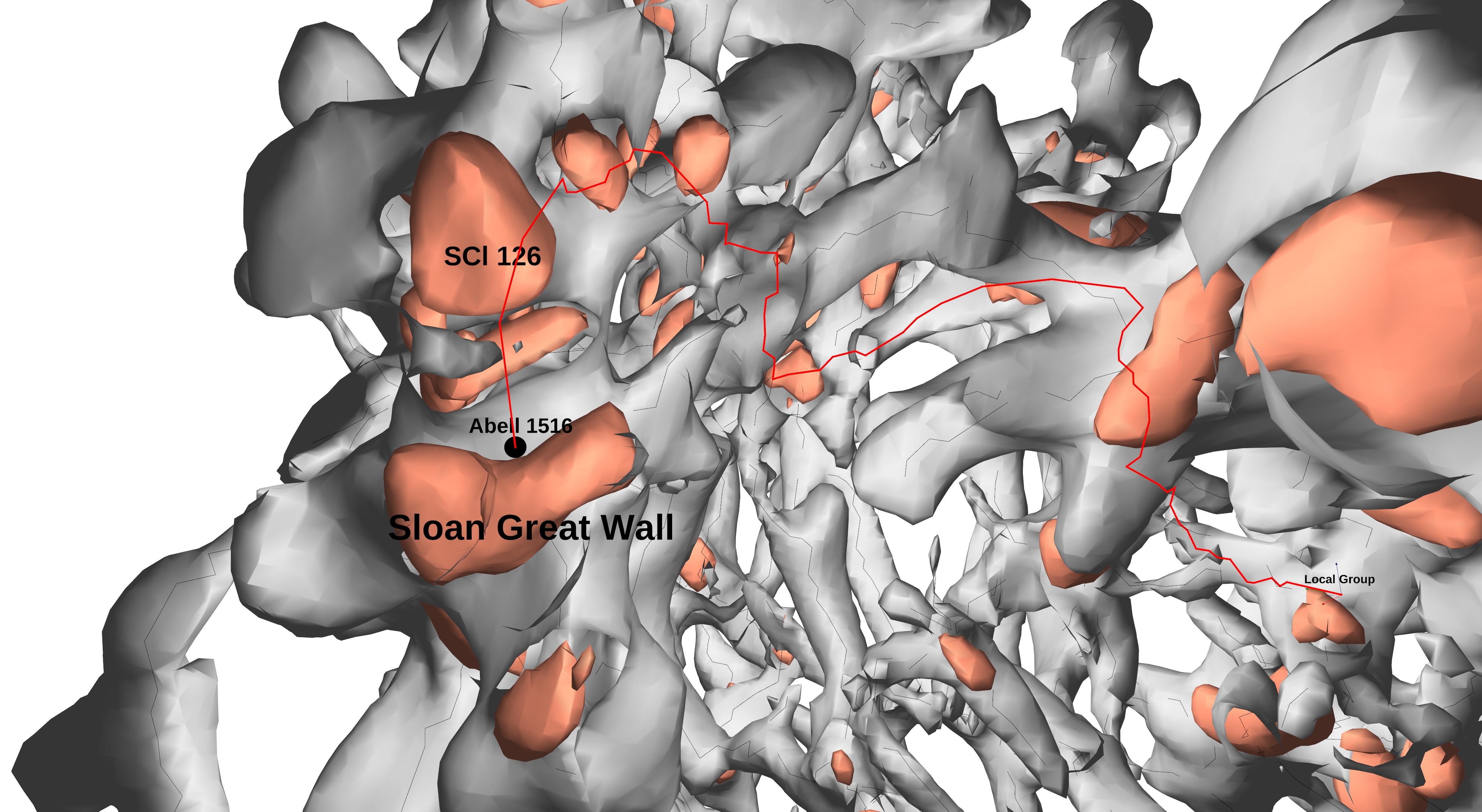}
    \caption{Aurélien's Path (red) along the V-web from the Milky Way to the Sloan BoA is explored in this \href{https://vimeo.com/1067350255/65d9e8bb92}{video}. Notable features along this path are the CfA Great Wall, Hercules Supercluster, and Boötes Supercluster.}
    \label{fig:path_aurelien}
\end{figure}

\begin{figure}
    \centering
\includegraphics[width=\linewidth]{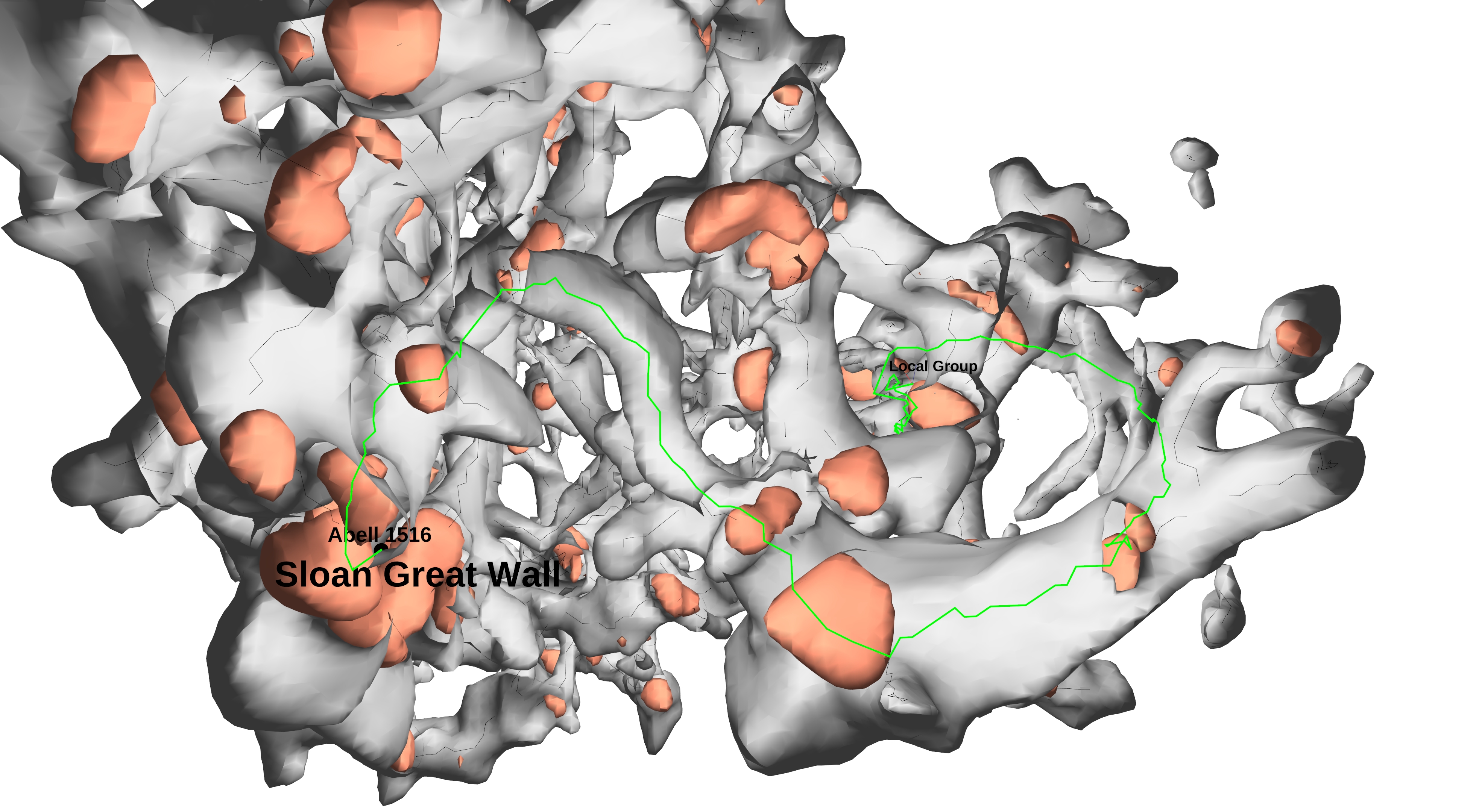}
    \caption{Noam's Path (green) along the V-web from the Milky Way to the Sloan BoA is explored in this \href{https://vimeo.com/1068924007/650afa6ca3}{video}. This is the southernmost path of all pathes. After a long detour through the galactic south across the Cen-Pup-PP filament, Perseus, Sculptor, Grus, the path crosses the Zone of Avoidance at the Circinus Bridge between the Triangulum Australis cluster and the Shapley Concentration. The path then warps around the Sloan Great Void through Boötes Supercluster.}
    \label{fig:path_noam}
\end{figure}

\begin{figure}
    \centering
\includegraphics[width=\linewidth]{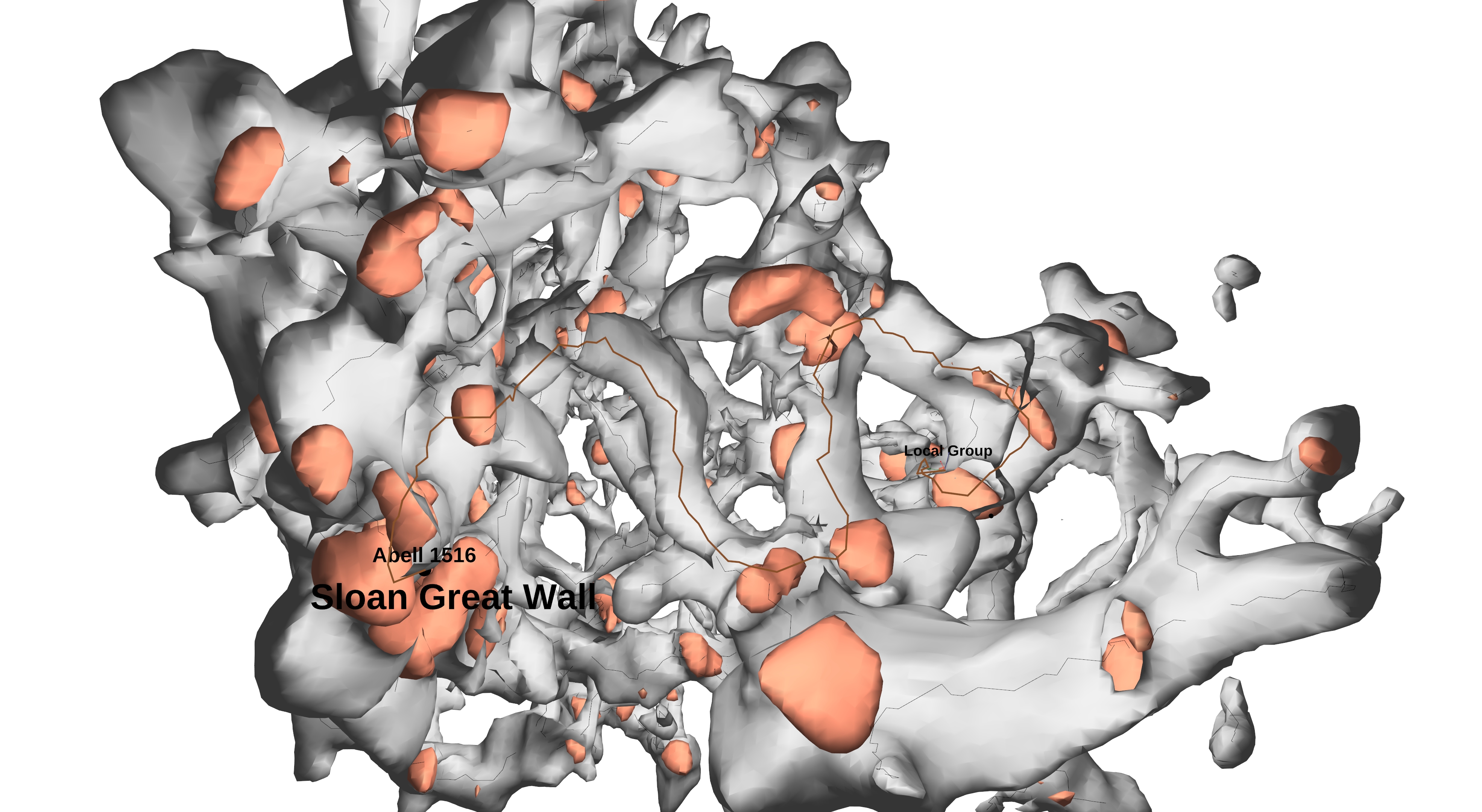}
    \caption{Yehuda's Path (brown) along the V-web from the Milky Way to the Sloan BoA is explored in this \href{https://vimeo.com/1069175251/d3c4fdabad}{video}. The path runs through the Great Attractor region to detour through the galactic south at Norma and the Norma-Pavo-Indus filament, returning to galactic north through the Scutum Bridge across the Zone of Avoidance in the vicinity of Ophiuchus cluster, onward to Hercules Supercluster, takes a lengthy road passing by the Shapley Concentration, to join with Noam's Path.}
    \label{fig:path_yehuda}
\end{figure}

\begin{figure}
    \centering
\includegraphics[width=\linewidth]{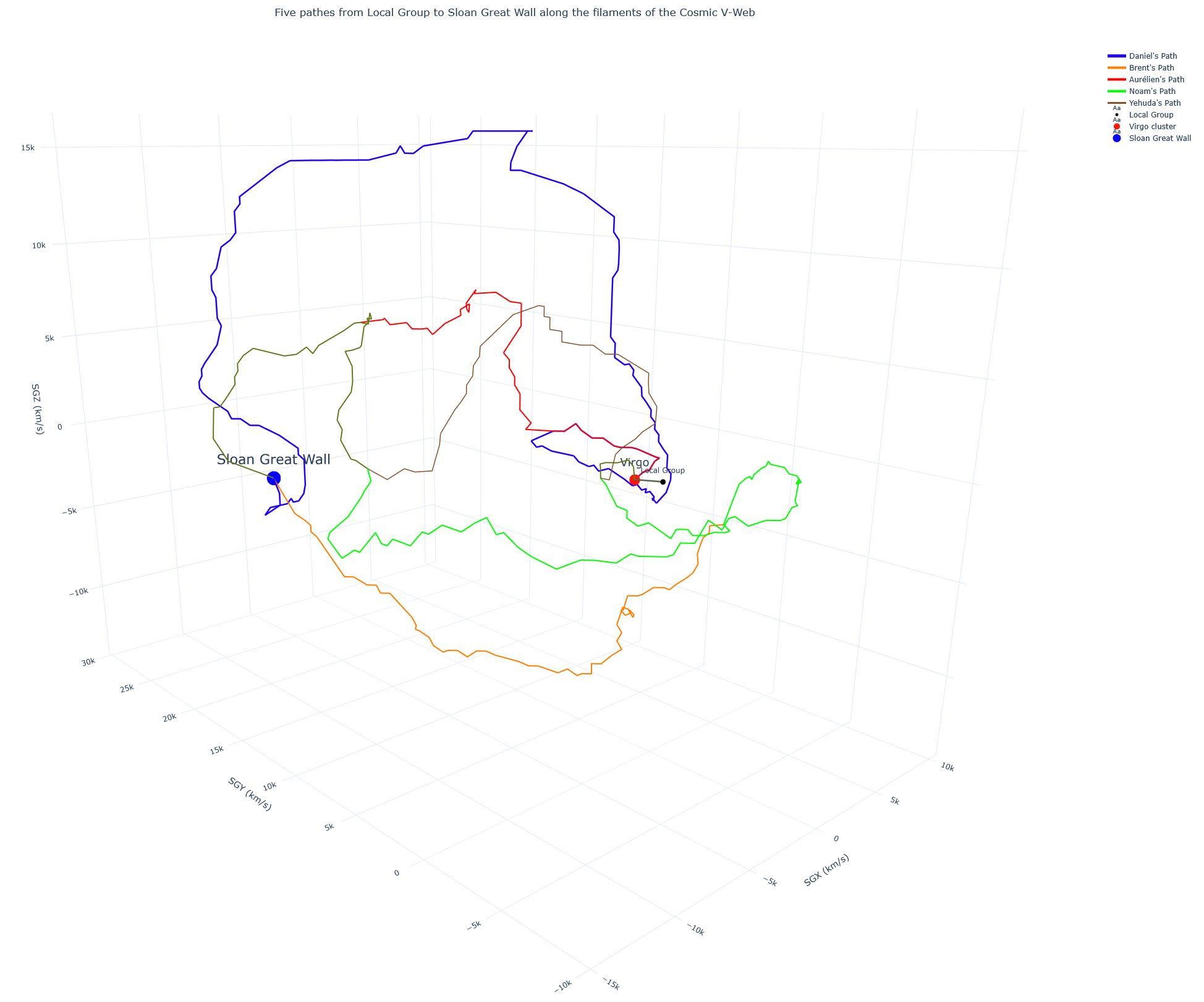}
    \caption{The skeletons of the five paths. View them in 3D in this \href{https://irfu.cea.fr/Projets/COAST/CF4_V-Web_pathes2Sloan.html}{interactive visualization}.}
    \label{fig:5pathes}
\end{figure}

The web of filaments is confusing even in the relatively small fraction of the Universe that is being explored in this study.
A taste of the complexity is provided by the five videos included in this appendix.
Each one follows the filament network from the Milky Way to the Sloan BoA in a different way.
The five paths are far from the only possibilities.
The paths are personalized as Daniel's Path (Fig.~\ref{fig:path_daniel}), Brent's Path (Fig.~\ref{fig:path_brent}), Aurélien's Path (Fig.~\ref{fig:path_aurelien}), Noam's Path (Fig.~\ref{fig:path_noam}), and Yehuda's Path (Fig.~\ref{fig:path_yehuda}).
Readers can access the V-web construction and find their own paths.
The skeletons of the five paths are shown together in the accompanying interactive model (Fig.~\ref{fig:5pathes}). 
It is to be emphasized that there is nothing special about  these paths.  They serve to visualize the connectedness of the V-web; easily obscured in the full V-web entanglement.

%%%%%%%%%%%%%%%%%%%%%%%%%%%%%%%%%%%%%%%

%%%%%%%%%%%%%%%%%%%%%%%%%%%%%%%%%%%%%%

\end{document}